\documentclass[aps,twocolumn,floatfix,superscriptaddress,showpacs,
]{revtex4-1}
\usepackage[utf8]{inputenc} 
\usepackage{todonotes}

\usepackage{graphicx}
\usepackage{dcolumn}
\usepackage{braket}
\usepackage{amsfonts}
\usepackage{amsmath}
\usepackage{amssymb}
\usepackage{amstext}
\usepackage{esint}

\usepackage{braket}
\usepackage{physics}

\usepackage{color,soul}
\usepackage{wasysym}
\usepackage{mathrsfs}
\usepackage{times}
\usepackage{hyperref}

\usepackage{float}
\usepackage[english]{babel}

\usepackage{bbm}
\usepackage{epstopdf} 
\usepackage[caption=false]{subfig}

\newif\ifContLineOne
\newif\ifContLineTwo
\newif\ifContLineThree
\def\conC#1{\vbox{\ialign{##\crcr
  \ifContLineThree\hrulefill\else\vphantom{\hrulefill}\fi\crcr
  \noalign{\kern3.2pt\nointerlineskip}
  \ifContLineTwo\hrulefill\else\vphantom{\hrulefill}\fi\crcr
  \noalign{\kern3.2pt\nointerlineskip}
  \ifContLineOne\hrulefill\else\vphantom{\hrulefill}\fi\crcr
  \noalign{\nointerlineskip}
  $\hfil\textstyle{\vbox to 14pt{}#1}\hfil$\crcr}}}
\def\DrawLeg#1#2{
  \kern-.2pt              
  \dimen2 =#1             
  \advance\dimen2 by 2pt  
  \dimen3 = 10.6pt        
  \dimen4 =3.6pt          
  \advance\dimen3 by -\dimen2 
  \multiply\dimen4 by #2
  \advance\dimen3 by \dimen4
  \raise\dimen2 \hbox{\vrule height\dimen3 width .4pt} 
  \kern-.2pt}             
\def\begC#1#2{\setbox0 =\hbox{$\textstyle{#2}$}
  \dimen0=.5\wd0 \dimen1=\ht0
  \conC{\hskip\dimen0}
  \count255=#1
  \ifnum\count255 =1 \ContLineOnetrue\else
  \ifnum\count255 =2 \ContLineTwotrue\else
  \ifnum\count255 =3 \ContLineThreetrue\fi\fi\fi
  \DrawLeg{\dimen1}{\count255}
  \conC{\hskip\dimen0}
  \kern-\dimen0\kern-\dimen0 \box0}
\def\endC#1#2{\setbox0 =\hbox{$\textstyle{#2}$}
  \dimen0=.5\wd0 \dimen1=\ht0
  \conC{\hskip\dimen0}
  \count255=#1
  \ifnum\count255 =1 \ContLineOnefalse\else
  \ifnum\count255 =2 \ContLineTwofalse\else
  \ifnum\count255 =3 \ContLineThreefalse\fi\fi\fi
  \DrawLeg{\dimen1}{\count255}
  \conC{\hskip\dimen0}
  \kern-\dimen0\kern-\dimen0 \box0}

\hypersetup{
pdfstartview={FitH},
pdftitle={2017},    
colorlinks=true,    
linkcolor=blue, 
citecolor=blue,   
filecolor=blue, 
urlcolor=blue   
}

\begin{document}

\title{Lieb and hole-doped ferrimagnetism, spiral, resonating  valence-bond states, and phase separation in large-U  $AB_{2}$ Hubbard chains}

\author{V. M. Martinez Alvarez}
\affiliation{Departamento de F\'{\i}sica, Laborat\'orio de F\'{\i}sica Te\'orica e Computacional,
Universidade Federal de Pernambuco, Recife 50670-901, Pernambuco, Brazil}
\author{ M. D. Coutinho-Filho}
\affiliation{Departamento de F\'{\i}sica, Laborat\'orio de F\'{\i}sica Te\'orica e Computacional,
Universidade Federal de Pernambuco, Recife 50670-901, Pernambuco, Brazil}


\begin{abstract}
The ground state (GS) properties of the quasi-one-dimensional $AB_2$ Hubbard model are investigated taking the effects of charge and spin quantum fluctuations on equal footing. In the strong-coupling regime, we derive a low-energy Lagrangian suitable to describe the ferrimagnetic phase at half filling and the phases in the hole-doped regime. At half filling, a perturbative spin-wave analysis allows us to find the GS energy, sublattice magnetizations, and Lieb total spin per unit cell of the effective quantum Heisenberg model, in very good agreement with previous results. In the challenging hole doping regime away from half filling, we derive the corresponding $t\textrm{-}J$ Hamiltonian. Under the assumption that charge and spin quantum correlations are decoupled, the evolution of the second-order spin-wave modes in the doped regime unveils the occurrence of spatially modulated spin structures and the emergence of phase separation in the presence of resonating-valence-bond states. We also calculate the doping-dependent GS energy and total spin per unit cell, in which case it is shown that the spiral ferrimagnetic order collapses at a critical hole concentration. Notably, our analytical results in the doped regime are in very good agreement with density matrix renormalization group studies, where our assumption of spin-charge decoupling is numerically supported by the formation of charge-density waves in anti-phase with the modulation of the magnetic structure.
\end{abstract}

\maketitle

\section{\label{sec:Introduction}introduction}

Much attention has been given to quantum phase transitions~\cite{SubirSachdev,Continentino2001},
which are phenomena characterized by the change of the nature of the
ground state (GS) driven by a non-thermal parameter: pressure, magnetic
field, doping, Coulomb repulsion, or competitive interactions. In
this context, the study of quasi-one-dimensional (quasi-1D) compounds
with ferrimagnetic properties~\cite{CoutinhoFilho2008,Ivanov2009}
has attracted considerable theoretical and experimental interest because
of their unique physical properties and very rich phase diagrams.
In particular, the GS of quasi-1D quantum ferrimagnets with $AB_{2}$ or $ABB'$ unit cell topologies (diamond or trimer chains) described by the Heisenberg or Hubbard models~\cite{MontenegroFilho2005} exhibit unsaturated spontaneous magnetization, ferromagnetic and antiferromagnetic spin-wave modes, effect of quantum fluctuations, and field-dependent magnetization
plateaus, among several other features of interest.  

Of special interest is the topological origin of GS magnetic long-range order associated with  the unit cell structure of the lattice~\cite{MontenegroFilho2005,SilvestreHoffmann1985,Macedo1995,Tian1996,Alcaraz1997,Raposo1997,*Raposo1999,Sierra1999,Martin2005}.
These studies have been motivated and supported by exact solutions
and rigorous results~\cite{LiebMattis,Lieb1968,Lieb1989,Lieb1995hubbard,Tasaki1998,*Tasaki1998nagaoka,Tian2004};
in particular, at half filling, the total spin per unit cell obeys
Lieb-Mattis~\cite{LiebMattis} (Heisenberg model) or Lieb's theorem~\cite{Lieb1989}
(Hubbard model). On the other hand, it has been verified that the
ferrimagnetic GS of spin-$1/2$ Heisenberg and Hubbard$/$$t$-$J$ $AB_{2}$
chains, under the effect of frustration~\cite{Yamamoto2007,MontenegroFrustration,Hida2008,Shimokawa2012,Giamarchi2014}  or doping~\cite{Macedo1995,Sierra1999,Montenegro2006,Montenegro2014}, are strongly affected by quantum fluctuations that might cause
its destruction and the occurrence of new exotic phases:  spiral incommensurate (IC) spin structures, Nagaoka ($U\rightarrow\infty$) and resonating-valence-bond (RVB) states, phase separation (PS), and Luttinger-liquid behavior. These features can enhance the phenomenology in comparison with a linear chain, which is dominated by the nontrivial Luttinger-liquid behavior that  exhibits fractional excitations~\cite{Giamarchi2004,Essler2005}, emergent fractionalized particles~\cite{CARMELO_2018}, and fractional-exclusion statistic properties~\cite{Haldane_1991,*Wu_1994} in the spin-incoherent regime~\cite{Vitoriano_2018}. In addition, investigations of transport properties in $AB_{2}$ chains, and related structures, have also unveiled very interesting features~\cite{Lopes2011,*Lopes2014}.

On the experimental side, studies~\cite{Drillon1993,Belik2005,Matsuda2005PhysRevB}
of the magnetic properties of homometallic phosphate compounds of
the family $\mathrm{A_{3}Cu_{3}\left(PO_{4}\right)_{4}}$ ($\mathrm{A=Ca}$,
$\mathrm{Sr}$, $\mathrm{Pb}$) suggest that in these materials
the line of trimers formed by spin-$1/2$ $\mathrm{Cu^{+2}}$ ions
antiferromagnetically coupled do exhibit ferrimagnetism of topological
origin. Further, compounds  $\mathrm{Ca_{3}}\mathrm{M}_{3}(\mathrm{PO_{4}})_{4}(\mathrm{M}=\mathrm{Ni},\mathrm{Co})$ with a wave-like layer structure built by zigzag M-chains exhibit antiferromagnetic ordering ($\mathrm{M}=\mathrm{Ni}$) or paramagnetic behavior ($\mathrm{M}=\mathrm{Co}$)~\cite{Guo_2017}. On the other hand, bimetallic compounds, such as $\mathrm{CuMn\left(S_{2}C_{2}O_{2}\right)_{2}\cdot7.5H_{2}O}$~\cite{Verdaguer1984},
can be modeled~\cite{Verdaguer1984,Tenorio2011,Strecka2016,*[See also: ][]Yan2015} by alternate
spin-$1/2$ - spin-$5/2$ chains and support interesting field-induced
quantum critical points and Luttinger-liquid phase~\cite{Tenorio2011}. In addition, frustrated
diamond ($AB_{2}$ topology) chains can properly model the compound
azurite, $\mathrm{Cu_{3}(CO_{3})_{2}(OH)_{2}}$, in which case the
occurrence of the $1/3$ magnetization plateau is verified at high fields~\cite{KikuchiPhysRevLett}
in agreement with topological arguments~\cite{Oshikawa1997} akin
to those invoked in the quantum Hall effect. The spin-1/2 trimer chain
compound $\mathrm{Cu_{3}(P_{2}O_{6}OH)_{2}}$, with antiferromagnetic
interactions only, also display the 1/3 magnetization plateau~\cite{Hase2006}.
Interestingly, it has been established that in azurite the magnetization
plateau is a dimer-monomer state~\cite{Rule2008}, i.e., the chain
is formed by pairs of $S=1/2$ monomers and  $S=0$ dimers, with a small local polarization of the diamond spins~\cite{Aimo_2009}, in agreement with density functional theory~\cite{Jeschke_2011}. These
dimer-monomer states have been found previously in the context of
modeling frustrated $AB_{2}$ chains~\cite{TakanoGSE,Okamoto1999,Okamoto2003},
 and confirmed through a modeling using quantum rotors~\cite{Tenoriorotors2009}. In contrast to azurite, whose dimers appear perpendicular to the chain direction, in the spin-1/2 inequilateral diamond-chain compounds~\cite{Morita_2017} $\mathrm{A}_{3}\mathrm{Cu}_{3}\mathrm{AlO}_{2}(\mathrm{SO}_{4})_{4}(\mathrm{A}=\mathrm{K},\mathrm{Rb},\mathrm{Cs})$,
the magnetic exchange interactions force the dimers to lie along
the sides of the diamond cells and the monomers form a 1D Heisenberg
chain. In fact, the low-energy excitations of these new compounds
have been probed and a Tomonaga-Luttinger spin liquid behavior identified~\cite{fujihala_2017_possible}. It is worth mentioning that strongly frustrated $AB_{2}$ chains can exhibit ladder-chain decoupling~\cite{MontenegroFrustration}, in which case the ladder is formed via the coupling between dimer
spins in neighboring $AB_{2}$ unit cells.

On the other hand, besides the above-mentioned quasi-1D compounds and related magnetic properties, considerable efforts have been devoted to the study of superconductivity and intriguing magnetic/charge ordered phases in doped materials~\cite{Dagotto_2005,Han_2015}, in particular the formation of spin-gapped states in compounds such as the family of doped $\mathrm{(La,Sr,Ca)_{14}Cu_{24}O_{41}}$. This compound is formed by one-dimensional $\mathrm{CuO_{2}}$ diamond chains, $\mathrm{(Sr, Ca)}$ layers, and two-leg $\mathrm{Cu_2O_3}$ ladders~\cite{Dagotto1999}. These results certainly stimulate experimental and theoretical investigations of quasi-1D compounds in the hole-doped regime, which is the main focus of our work, as described in the following.

In this work, we shall employ an analytical approach suitable to describe
the strongly coupled Hubbard model on doped  $AB_{2}$ chains, which
were the object of recent numerical studies through density matrix renormalization group (DMRG) techniques~\cite{Montenegro2014}.
Our functional integral approach, combined with a perturbative expansion
in the strong-coupling regime, was originally proposed to study the
doped Hubbard chain~\cite{Weng_1991prl,*Weng_1992prb}, and later
adapted to describe various doped-induced phase transitions in the
$U=\infty$ $AB_{2}$ Hubbard chain~\cite{Oliveira_2009}. In addition, this approach was used to describe the doped strongly coupled Hubbard model on the honeycomb lattice~\cite{Ribeiro2015}, whose results are very rewarding, particularly those for the GS energy and magnetization in the doped regime, which compare very well with Grassmann tensor product numerical studies~\cite{Gu2014Wen}.

The paper is organized as follows: in Sec.~\ref{sec:Functional}
we review the functional integral representation of the Hubbard Hamiltonian
in terms of Grassmann fields (charge degrees of freedom) and spin
$SU(2)$ gauge fields (spin degrees of freedom). In Sec.~\ref{sec:Diagonal}
we diagonalize the Hamiltonian associated with the charge degree of
freedom and obtain a perturbative low-energy theory suitable to describe
the ferrimagnetic phase at half filling and the phases in the hole-doped regimes. In Sec.~\ref{sec:HALF-FILLING}, we show that the resultant
Hamiltonian at half filling and large-U maps onto the spin-$1/2$
quantum Heisenberg model. In this regime, a perturbative series expansions
in powers of $1/S$ of the spin-wave modes is presented, which allows us to calculate the GS energy, sublattice
magnetizations, and Lieb GS total spin per unit cell in very good agreement with previous estimates. In
Sec.~\ref{sec:t-J}, we derive the low-energy effective $t$-$J$
Hamiltonian, which accounts for both charge and spin quantum fluctuations. We also present the  evolution of  the second-order spin-wave modes, GS energy and total spin per unit cell under hole doping, thus identifying the occurrence of spatially modulated spin structures, with non-zero and zero GS total spin, and phase separation involving the later spin structure and RVB states at hole concentration $1/3$. Remarkably, these predictions are in very good agreement with the DMRG data reported in Ref.~[\onlinecite{Montenegro2014}]. Lastly, 
in Sec.~\ref{sec:CONCLUSIONS}, we present a summary and concluding remarks concerning the reported results.

\section{\label{sec:Functional}Functional-Integral representation}

The Hamiltonian of the one-band Hubbard model on chains with $AB_{2}$
unit cell topology is given by~\cite{Macedo1995,Tian1996,Raposo1997,Raposo1999}:\begin{equation}
\mathcal{H}=-\sum_{\left\langle i\alpha,j\beta\right\rangle \sigma}\{t_{ij}^{\alpha\beta}\hat{c}_{i\alpha\sigma}^{\dagger}\hat{c}_{j\beta\sigma}+\textrm{H.c.}\}+U\sum_{i\alpha}\hat{n}_{i\alpha\uparrow}\hat{n}_{i\alpha\downarrow},\label{Hamil}
\end{equation}where $i=1,\ldots,N_{c}\,(=N/3)$ is the specific position of the
unit cell, whose length is set to unity, $N_{c}$ ($N$) is the number of cells (sites), $\alpha,\beta=A,\,B_{1},\,B_{2}$
denote the type of site within the unit cell, $\hat{c}_{i\alpha\sigma}^{\dagger}$
($\hat{c}_{i\alpha\sigma}$) is the creation (annihilation) operator
of electrons with spin $\sigma$ ($=\uparrow,\downarrow$) at site
$\alpha$ of cell $i$, and $\hat{n}_{i\alpha\sigma}=\hat{c}_{i\alpha\sigma}^{\dagger}\hat{c}_{i\alpha\sigma}$
is the occupancy number operator. The first term in Eq.~(\ref{Hamil})
describes electron hopping, with energy $t_{ij}^{\alpha\beta}\equiv t$,
allowed only between nearest neighbors $A\textrm{-}B_{1}$ and $A\textrm{-}B_{2}$
linked sites of sublattices $A$ and $B$ (bipartite lattice), and
the second one is the on-site Coulombian repulsive interaction $U>0$, which contributes only in the case of double occupancy of the site $i\alpha$.

At this point, it is instructive to digress on some fundamental aspects of the formalism used in our work~\cite{Weng_1991prl,Weng_1992prb,Oliveira_2009,Ribeiro2015}. With regard to the large-U doped Hubbard chain~\cite{Weng_1991prl,Weng_1992prb},
$U=\infty$ $AB_{2}$ Hubbard chain~\cite{Oliveira_2009} and the
Hubbard model on the honeycomb lattice~\cite{Ribeiro2015}, it has
been shown that the particle density product in Eq.~(\ref{Hamil})
can be treated through the use of a decomposition procedure, which
consists in expressing $\hat{n}_{i\alpha\uparrow}\hat{n}_{i\alpha\downarrow}$
in terms of charge and spin operators:\begin{equation}
\hat{n}_{i\alpha\uparrow}\hat{n}_{i\alpha\downarrow}=\frac{1}{2}\hat{\rho}_{i\alpha}-2(\hat{\mathbf{S}}_{i\alpha}\cdot\mathbf{n}_{i\alpha})^{2},\label{dec1}
\end{equation}where
\begin{equation}\hat{\mathbf{S}}_{i\alpha}=1/2\sum_{\sigma\sigma^{'}}\hat{c}_{i\alpha\sigma^{'}}^{\dagger}\boldsymbol{\sigma}_{\sigma^{'}\sigma}\hat{c}_{i\alpha\sigma},\label{equation3}\end{equation}and\begin{equation}\hat{\rho}_{i\alpha}=\hat{n}_{i\alpha\uparrow}+\hat{n}_{i\alpha\downarrow},\label{rho4}\end{equation}are the spin-1/2 and charge-density operators, respectively, $\mathbf{\sigma}_{\sigma^{'}\sigma}$ denotes the Pauli matrix elements $(\hbar\equiv1)$, and $\mathbf{n}_{i\alpha}$ is   an arbitrary unit vector. In fact, Eq.~(\ref{dec1}) follows from the identity:  $\frac{1}{2}\hat{\rho}_{i\alpha}-\hat{n}_{i\alpha\uparrow}\hat{n}_{i\alpha\downarrow}=2(\hat{S}_{i\alpha}^{x,y,z})^{2}=2(\hat{\mathbf{S}}_{i\alpha}\cdot\mathbf{n}_{i\alpha})^{2}$.  The convenience of using the decomposition
defined in Eq.~(\ref{dec1}), with explicit spin-rotational invariance for the large-U Hubbard model, was discussed
at length in Refs.~[\onlinecite{Weng_1991prl,Weng_1992prb,Oliveira_2009,Ribeiro2015}].

We start by using the Trotter-Suzuki formula~\cite{fradkin2013field,negele1988}, 
which allows us to write the partition function, $\mathcal{Z}=\textrm{Tr}\left[\exp(-\beta\mathcal{H})\right]$,
at a temperature $k_{B}T\equiv1/\beta$, as $\mathcal{Z}=\textrm{Tr}\{\hat{T}\prod_{r=1}^{M}\exp[-\delta\tau\mathcal{H}(\tau_{r})]\}$,
where $\hat{T}$ denotes the time-ordering operator, the total imaginary
time interval is formally sliced into $M$ discrete intervals of equal
size $\delta\tau=\tau_{r}-\tau_{r-1}$, $r=1,2,...,M,$ with $\tau_{0}=0,$ and
$\tau_{M}=\beta=M\delta\tau$, under the limits $M\rightarrow\infty$
and $\delta\tau\rightarrow0$. We shall now introduce, between each discrete time interval, an overcomplete basis of fermionic coherent states~\cite{fradkin2013field,negele1988},
$1=\int\prod_{i\alpha\sigma}dc_{i\alpha\sigma}^{\dagger}dc_{i\alpha\sigma}\exp(-\sum_{i\alpha\sigma}c_{i\alpha\sigma}^{\dagger}c_{i\alpha\sigma})\arrowvert\{c_{i\alpha\sigma}\}\rangle\langle\{c_{i\alpha\sigma}\}\arrowvert$,
where $\{c_{i\alpha\sigma}^{\dagger},c_{i\alpha\sigma}$\} denotes
a set of Grassmann fields satisfying anti-periodic boundary conditions: $c_{i\alpha\sigma}^{\dagger}(0)=-c_{i\alpha\sigma}^{\dagger}(\beta)$
and $c_{i\alpha\sigma}(0)=-c_{i\alpha\sigma}(\beta)$; while the set of unit vectors defines the vector field $\{\mathbf{n}_{i\alpha}\}$,   satisfying periodic ones: $\mathbf{n}_{i\alpha}(0)=\mathbf{n}_{i\alpha}(\beta)$, under a weight functional (see below). Thereby, following standard procedure~\cite{fradkin2013field,negele1988}, the partition function reads:\begin{equation}\mathcal{Z}=\int\prod_{i\alpha\sigma}\mathcal{D}c_{i\alpha\sigma}^{\dagger}\mathcal{D}c_{i\alpha\sigma}\prod_{i\alpha}\mathcal{D}^{2}\mathbf{n}_{i\alpha}W(\{\mathbf{n}_{i\alpha}\})e^{-\int_{0}^{\beta}\mathcal{L}(\tau)d\tau},\label{partlim}
\end{equation}where the pertinent measures are defined by\begin{equation}
\mathcal{D}c_{i\alpha\sigma}^{\dagger}\mathcal{D}c_{i\alpha\sigma}\equiv\lim_{M\rightarrow\infty, \delta\tau\rightarrow0}\prod_{r=1}^{M-1}dc_{i\alpha\sigma}^{\dagger}(\tau_{r})dc_{i\alpha\sigma}(\tau_{r}),\label{Dc}\end{equation}\begin{equation}\mathcal{D}^{2}\mathbf{n}_{i\alpha}\equiv\lim_{M\rightarrow\infty, \delta\tau\rightarrow0}\prod_{r=1}^{M-1}d^{2}\mathbf{n}_{i\alpha}(\tau_{r}),\label{Dn}\end{equation}the weight functional, $W(\{\mathbf{n}_{i\alpha}\})$, satisfies a normalization condition at each discrete imaginary time $\tau_r$:\begin{equation}\int\prod_{i\alpha}d^{2}\mathbf{n}_{i\alpha}W(\{\mathbf{n}_{i\alpha}(\tau_r)\})=1,\label{Wei}\end{equation}and the Lagrangian density $\mathcal{L}(\tau)$ is written in the form:
\begin{align}
\mathcal{L}(\tau) & =\sum_{i\alpha\sigma}c_{i\alpha\sigma}^{\dagger}\partial_{\tau}c_{i\alpha\sigma}-\sum_{ij\alpha\beta\sigma}(t_{ij}^{\alpha\beta}c_{i\alpha\sigma}^{\dagger}c_{j\beta\sigma}+\textrm{H.c.})\nonumber \\
 & +U\sum_{i\alpha}[\frac{\rho_{i\alpha}}{2}-2(\mathbf{S}_{i\alpha}\cdot\mathbf{n}_{i\alpha})^2].\label{eq:Lagrangiana1}
\end{align}

In order to fix $W(\{\mathbf{n}_{i\alpha}\})$ one should notice that, in the operator formalism: $\hat{\rho}_{i\alpha}^{2}=\hat{\rho}_{i\alpha}+2\hat{n}_{i\alpha\uparrow}\hat{n}_{i\alpha\downarrow}$. Therefore, using Eq.~(\ref{dec1}), the following identity holds~\cite{Weng_1991prl,Weng_1992prb}:\begin{equation}
2(\mathbf{\hat{S}}_{i\alpha}\cdot\mathbf{n}_{i\alpha})^{2}=\frac{\hat{\rho}_{i\alpha}(2-\hat{\rho}_{i\alpha})}{2},\label{eq3}
\end{equation}which means that  the square of the spin component operator along the $\mathbf{n}_{i\alpha}$ direction has zero eigenvalues if the site is vacant or doubly occupied, and a nonzero value only for singly occupied sites, i.e.,  ${(\mathbf{\hat{S}}_{i\alpha}\cdot\mathbf{n}_{i\alpha})^{2}=1/4}$. Now, taking advantage of the choice of $\mathbf{n}_{i\alpha}$, the local spin-polarization and spin-quantization axes are both chosen along the $\mathbf{n}_{i\alpha}$ direction. Therefore, for singly occupied sites, we find $\mathbf{S}_{i\alpha}\cdot\mathbf{n}_{i\alpha}=p_{i\alpha}/2$, with $p_{i\alpha}=\pm 1$, corresponding to the two possible spin-1/2 states. Further, by incorporating vacancy and double occupancy possibilities, corresponding to the four possible local states of the Hubbard model, one can write~\cite{Weng_1991prl,Weng_1992prb}
\begin{equation}
p_{i\alpha}\mathbf{\hat{S}}_{i\alpha}\cdot\mathbf{n}_{i\alpha}=\frac{\hat{\rho}_{i\alpha}(2-\hat{\rho}_{i\alpha})}{2},\label{eq4}
\end{equation}
with $p_{i\alpha}^2=(\pm1)^2$. We stress that, due to fermion operator properties, the square of Eq.~(\ref{eq4}) reproduces Eq.~(\ref{eq3}), and a comparison between them implies, at arbitrary doping and U value, the formal equivalence between $2(\hat{\mathbf{S}}_{i\alpha}\cdot\mathbf{n}_{i\alpha})^{2} $ and $p_{i\alpha}(\hat{\mathbf{S}}_{i\alpha}\cdot\mathbf{n}_{i\alpha})$. In this context, we remark that the original Coulomb repulsion term of the Hubbard Hamiltonian in Eq.~(\ref{Hamil}) is formally and energetically (eigenvalues) equivalent to both that in Eq.~(\ref{eq:Lagrangiana1}) or in its linear version through the following replacement: $2(\hat{\mathbf{S}}_{i\alpha}\cdot\mathbf{n}_{i\alpha})^{2}\rightarrow p_{i\alpha}(\hat{\mathbf{S}}_{i\alpha}\cdot\mathbf{n}_{i\alpha})$. Indeed, using the constraint in Eq.~(\ref{eq4}) we find,
$U\sum_{i\alpha}[\frac{\rho_{i\alpha}}{2}-p_{i\alpha}(\mathbf{S}_{i\alpha}\cdot\mathbf{n}_{i\alpha})]=U\sum_{i\alpha}[\frac{\rho_{i\alpha}}{2}-\frac{1}{2}\rho_{i\alpha}(2-\rho_{i\alpha})]$,  which is zero for $\rho_{i\alpha}=0,1$; whereas, as expected, for double occupied sites, $\rho_{i\alpha}=2$, the local energy is $U$. 
Therefore, Eq.~(\ref{eq4}) in its Grassmann version, can be enforced by a proper choice of the normalized weight functional:\begin{align}W & (\{\mathbf{n}_{i\alpha}\})=\lim_{M\rightarrow\infty,\delta\tau\rightarrow0}\prod_{r=1}^{M}W(\{\mathbf{n}_{i\alpha}(\tau_{r})\})\nonumber\\ & =\mathcal{C}\exp\Big\{-\int_{0}^{\beta}d\tau\gamma\sum_{i\alpha}[p_{i\alpha}\mathbf{S}_{i\alpha}\cdot\mathbf{n}_{i\alpha}-\frac{\rho_{i\alpha}}{2}(2-\rho_{i\alpha})]^{2}\Big\},\label{eq9}
\end{align}where $\gamma\rightarrow\infty$ in the continuum limit ($M\rightarrow\infty$, $\delta\tau\rightarrow0$), with delta-function peaks at the four local states of the Hubbard model, and $\mathcal{C}$ is a normalization factor such that Eq.~(\ref{Wei}) holds. In fact, the product of $W(\{\mathbf{n}_{i\alpha}(\tau_r)\})$ in Eq.~(\ref{eq9}) generates a sum in $r$ in the exponential of the suitable chosen Gaussian function, i.e., $W(\{\mathbf{n}_{i\alpha}\})$ is such that in the continuum limit, $M\rightarrow\infty,\delta\tau\rightarrow0$, Eq.~(\ref{eq9}) obtains with a diverging $\gamma$, as pointed out in Ref.~[\onlinecite{Weng_1992prb}]. In this way, using Eq.~(\ref{eq9})  for the weight functional
in Eq.~(\ref{partlim}) for the partition function $\mathcal{Z}$, and integrating over $\{\mathbf{n}_{i\alpha}\}$, the Lagrangian density $\mathcal{L}(\tau)$ in Eq. (\ref{eq:Lagrangiana1}) can thus be written in the following linearized form~\cite{Weng_1991prl}:
\begin{align}
\mathcal{L}(\tau) & =\sum_{i\alpha\sigma}c_{i\alpha\sigma}^{\dagger}\partial_{\tau}c_{i\alpha\sigma}-\sum_{ij\alpha\beta\sigma}(t_{ij}^{\alpha\beta}c_{i\alpha\sigma}^{\dagger}c_{j\beta\sigma}+\textrm{H.c.})\nonumber \\
 & +U\sum_{i\alpha}[\frac{\rho_{i\alpha}}{2}-p_{i\alpha}(\mathbf{S}_{i\alpha}\cdot\mathbf{n}_{i\alpha})],\label{eq:Lagrangiana10}
\end{align}
where the constraint in Eq. (\ref{eq4}) was explicitly used.

Now, since we are interested in studying the GS properties of the $AB_{2}$ Hubbard chains, we choose the staggered factor $p_{i\alpha}=+1\,\,(-1)$ at sites $\alpha=B_{1},\,B_{2}$ $(A)$, consistent with the long-range ferrimagnetic GS predicted by Lieb's theorem at half filling and for any $U$ value~\cite{Lieb1989,Macedo1995,Tian1996}, in which case we assume broken rotational symmetry along the
$z$-axis. In this context, by considering the symmetry exhibited by the ferrimagnetic order,
let us define the $SU(2)/U(1)$ unitary rotation matrix~\cite{tung1985group} 
\begin{equation}
U_{i\alpha}=\left[\begin{array}{cc}
\cos\left(\frac{{\textstyle \theta_{i\alpha}}}{{\textstyle 2}}\right) & -\sin\left(\frac{{\textstyle \theta_{i\alpha}}}{{\textstyle 2}}\right)e^{-i\phi_{i\alpha}}\\
\sin\left(\frac{{\textstyle \theta_{i\alpha}}}{{\textstyle 2}}\right)e^{i\phi_{i\alpha}} & \cos\left(\frac{{\textstyle \theta_{i\alpha}}}{{\textstyle 2}}\right)
\end{array}\right],\label{Ufor}
\end{equation}
where $\theta_{i\alpha}$ is the polar angle between the $z$-axis
and the unit local vector $\mathbf{n}_{i\alpha}$ and $\phi_{i\alpha}\in[0,2\pi)$
is an arbitrary azimuth angle due to the $U(1)$ gauge freedom of
choice for $U_{i\alpha}$. Moreover, a new set of Grassmann fields, \{$a_{i\alpha\sigma}^{\dagger},a_{i\alpha\sigma}$\}
can be obtained, according to the transformation:
\begin{equation}
c_{i\alpha\sigma}=\sum_{\sigma'}(U_{i\alpha})_{\sigma\sigma'}a_{i\alpha\sigma'},\label{newfield}
\end{equation}
that locally rotates each unit vector $\mathbf{n}_{i\alpha}$ to the $z$-direction. On the other hand, if we express the product $\boldsymbol{\sigma}\cdot\mathbf{n}_{i\alpha}$ in matrix form:\begin{equation}
\boldsymbol{\sigma}\cdot\mathbf{n}_{i\alpha}=\left[\begin{array}{cc}
\cos\left(\theta_{i\alpha}\right) & \sin\left(\theta_{i\alpha}\right)e^{-i\phi_{i\alpha}}\\
\sin\left(\theta_{i\alpha}\right)e^{i\phi_{i\alpha}} & -\cos\left(\theta_{i\alpha}\right)
\end{array}\right],
\end{equation}
we obtain, after using Eq.~(\ref{Ufor}),\begin{equation}
U_{i\alpha}^{\dagger}(\boldsymbol{\sigma}\cdot\mathbf{n}_{i\alpha})U_{i\alpha}=\sigma^{z},\label{UsigmazU}
\end{equation}which explicitly manifest the broken rotational symmetry along the $z$-axis. In this way, by substituting  Eqs.~(\ref{Ufor}) and (\ref{newfield})  into Eq.~(\ref{equation3}), and using the above result, we find\begin{align}
\mathbf{S}_{i\alpha}\cdot\mathbf{n}_{i\alpha} & =\frac{1}{2}\sum_{\sigma\sigma^{'}}a_{i\alpha\sigma}^{\dagger}[U_{i\alpha}^{\dagger}(\boldsymbol{\sigma}\cdot\mathbf{n}_{i\alpha})U_{i\alpha}]_{\sigma\sigma^{'}}a_{i\alpha\sigma^{'}}\nonumber \\
 & =\frac{1}{2}\sum_{\sigma\sigma^{'}}a_{i\alpha\sigma}^{\dagger}(\sigma_{z})_{\sigma\sigma^{'}}a_{i\alpha\sigma^{'}}\equiv S_{i\alpha}^{z};\label{eq:esez}
\end{align}thereby, the constraint in Eq.~(\ref{eq4}) can be written in the form\begin{equation}
\mathbf{S}_{i\alpha}\cdot\mathbf{n}_{i\alpha}=p_{i\alpha}\frac{\rho_{i\alpha}(2-\rho_{i\alpha})}{2}=\frac{1}{2}(a_{i\alpha\uparrow}^{\dagger}a_{i\alpha\uparrow}-a_{i\alpha\downarrow}^{\dagger}a_{i\alpha\downarrow}),\label{stilda}
\end{equation}where $p_{i\alpha}=+1\,\,(-1)$ at sites $\alpha=B_{1},\,B_{2}$ $(A)$. The choice of $p_{i\alpha}$
above implies Lieb's ferrimagnetic ordering with the set
$\{\theta_{iA}=\theta_{iB_{1}}=\theta_{iB_{2}}=0\}$, for all $i$, at half filling. However, in the hole doped regime away from half filling, the $\theta_{i\alpha}$'s can be nonzero (e.g., $\theta_{i\alpha}=\pi$ for a spin flip, leading to a change in the sign of $S_{i\alpha}^{z}$); further, $S_{i\alpha}^{z}$ can be zero either by the presence of holes or doubly occupied sites ($a_{i\alpha\uparrow}^{\dagger}a_{i\alpha\uparrow}=a_{i\alpha\downarrow}^{\dagger}a_{i\alpha\downarrow}$). Lastly, using Eqs.~(\ref{newfield}) and (\ref{stilda}) into the Lagrangian, Eq.~(\ref{eq:Lagrangiana10}), we find, after suitable rearrangement of terms,
\begin{equation}
\mathcal{L}(\tau)=\mathcal{L}_{0}(\tau)+\mathcal{L}_{n}(\tau),\label{eq:lagangiano general}
\end{equation}
where both Lagrangians are quadratic in the Grassmann fields:  \begin{align}
\mathcal{L}_{0}(\tau)  =&\sum_{i\alpha\sigma}a_{i\alpha\sigma}^{\dagger}\partial_{\tau}a_{i\alpha\sigma}-\sum_{i\alpha j\beta\sigma}(t_{ij}^{\alpha\beta}a_{i\alpha\sigma}^{\dagger}a_{j\beta\sigma}+\textrm{H.c.})\nonumber \\
 & +\frac{U}{2}\sum_{i\alpha\sigma}(1-p_{i\alpha}\sigma)a_{i\alpha\sigma}^{\dagger}a_{i\alpha\sigma},\label{lag0}
\end{align}
and
\begin{align}
\mathcal{L}_{n} & (\tau)=\sum_{i\alpha\sigma\sigma'}a_{i\alpha\sigma'}^{\dagger}(U_{i\alpha}^{\dagger}\partial_{\tau}U_{i\alpha})_{\sigma'\sigma}a_{i\alpha\sigma}\nonumber \\
 & -\sum_{i\alpha j\beta\sigma\sigma'}t_{ij}^{\alpha\beta}[a_{i\alpha\sigma'}^{\dagger}(U_{i\alpha}^{\dagger}U_{j\beta}-1)_{\sigma'\sigma}a_{j\beta\sigma}+\textrm{H.c.}],\label{eq:LanN}
\end{align}
with the first term in both Eqs.~(\ref{lag0}) and (\ref{eq:LanN}) being originated from the first term in Eq.~(\ref{eq:Lagrangiana10}), the second ones come from the hopping term in Eq.~(\ref{eq:Lagrangiana10}), after a rearrangement of terms, while the last one in Eq.~(\ref{lag0}) (proportional to $U$) is obtained by using Eq.~(\ref{stilda}) in the last term of Eq.~(\ref{eq:Lagrangiana10}). It is worth mentioning that only charge  degrees of freedom (Grassmann fields) appear 
in $\mathcal{L}_{0}(\tau)$, and spin degrees of freedom under the constraint in Eq.~(\ref{stilda}) [$SU(2)$ gauge fields $\{U_{i\alpha}^{\dagger},U_{i\alpha}\}$, which carry all the information on the vector field $\{\mathbf{n}_{i\alpha}\}$]   are now restricted
to $\mathcal{L}_{n}(\tau)$, which includes both spin and charge degrees of freedom. 

In the large-U regime, double occupancy is energetically unfavorable and the factor $2-\rho_{i\alpha}$ is no longer needed in Eq.~(\ref{stilda}), i.e., $\mathbf{S}_{i\alpha}\cdot\mathbf{n}_{i\alpha}=p_{i\alpha}\frac{\rho_{i\alpha}}{2}$, with $\rho_{i\alpha}=0$ or $1$.  In this case, a proper perturbative analysis
will allow us to study hole doping effects in Sec.~\ref{sec:t-J} in a macroscopic fashion, so we define\begin{equation}
\delta=1-\frac{1}{N}\sum_{i\alpha}\left\langle \rho_{i\alpha}\right\rangle,\label{e23}
\end{equation}which measures the thermodynamic average of hole doping away from half filling. In this context (strong-coupling limit), we take advantage of results derived from $\mathcal{L}_{0}(\tau)$ (charge effects in Sec.~\ref{sec:Diagonal}), and at half filling (Sec.~\ref{sec:HALF-FILLING}),  in which case charge degrees of freedom are frozen.

\section{\label{sec:Diagonal}Charge degrees of freedom and the strong-coupling limit}
In this section, we shall first diagonalize the Hamiltonian associated with the Lagrangian $\mathcal{L}_{0}(\tau)$ through the use of a special symmetry property of the $AB_2$ chains and a canonical transformation in reciprocal space.
Then, by introducing a perturbative expansion in the strong-coupling
regime, a low-energy effective Lagrangian for the $AB_{2}$ Hubbard
chains at half filling and in the doped regime will be obtained.

\subsection{\label{Charge-degrees-of-freedom}Charge degrees of freedom}

We begin our discussion by considering the Lagrangian $\mathcal{L}_{0}$
in Eq.~(\ref{lag0}), and its corresponding Hamiltonian $\mathcal{H}_{0}$,
free of the $SU(2)$ gauge fields. By performing
the Legendre transformation: $\mathcal{H}_{0}=-\sum_{i\alpha\sigma}\frac{\partial\mathcal{L}_{0}}{\partial(\partial_{\tau}a_{i\alpha\sigma})}\partial_{\tau}a_{i\alpha\sigma}+\mathcal{L}_{0}$,
where $\frac{\partial\mathcal{L}_{0}}{\partial(\partial_{\tau}a_{i\alpha\sigma})}=a_{i\alpha\sigma}^{\dagger}$,
the resulting $\mathcal{H}_{0}$ is given by
\begin{align}
\mathcal{H}_{0}= & -\sum_{\left\langle i\alpha,j\beta\right\rangle \sigma}(t_{ij}^{\alpha\beta}a_{i\alpha\sigma}^{\dagger}a_{j\beta\sigma}+\textrm{H.c.})\nonumber \\
 & +\frac{U}{2}\sum_{i\alpha\sigma}(1-p_{i\alpha}\sigma)a_{i\alpha\sigma}^{\dagger}a_{i\alpha\sigma}.\label{eq:Hamil01}
\end{align}
Further, since  $\mathcal{H}_{0}$ ($\mathcal{L}_{0}$) is quadratic in the Grassmann fields, the solution for the energy of the system is given by  $\mathcal{H}_{0}$ in its diagonalized form~\cite{negele1988}.

The $AB_{2}$ unit cell topology exhibits a symmetry~\cite{Alcaraz1997,Sierra1999,Montenegro2006,Oliveira_2009,Montenegro2014}
under the exchange of the labels of the $B$ sites in a given unit
cell. Thus, we can construct a new set of Grassmann fields possessing
this symmetry, i.e., either symmetric or antisymmetric with respect
to the exchange operation $B_{1}\leftrightarrow B_{2}$:
\begin{equation}
(d_{i\sigma},e_{i\sigma})= \frac{1}{\sqrt{2}}(a_{iB_{1}\sigma}\pm a_{iB_{2}\sigma}), \,\,\,\,\, b_{i\sigma}= a_{iA\sigma}.\label{SymmetryProperty}
\end{equation}In addition, as a signature of the quasi-1D structure of the $AB_{2}$ chains, we notice that the $B_{1}$ and $B_{2}$ sites are located at a distance $1/2$ (in units of length) ahead of the $A$ site. Therefore, after Fourier transforming the above  Grassmann fields, i.e., $\{d_{i,\sigma},e_{i,\sigma},b_{i,\sigma}\}=\frac{1}{\sqrt{N_{c}}}\sum_{k}e^{ikx_{i}}\{d_{k,\sigma},e_{k,\sigma},b_{k,\sigma}\}$, it is convenient to  introduce a phase factor $e^{\frac{ik}{2}}$
through the following transformation~\cite{Oliveira_2009}: $(A_{k\sigma},B_{k\sigma})=\frac{1}{\sqrt{2}}(d_{k\sigma}\pm e^{\frac{ik}{2}}b_{k\sigma})$,
so that $\mathcal{H}_{0}$ in Eq.~(\ref{eq:Hamil01}) thus becomes
\begin{align}
 & \mathcal{H}_{0}=\sum_{k\sigma}\varepsilon_{k}[A_{k\sigma}^{\dagger}A_{k\sigma}-B_{k\sigma}^{\dagger}B_{k\sigma}]+\frac{U}{2}\sum_{k\sigma}(1-\sigma)e_{k\sigma}^{\dagger}e_{k\sigma}\nonumber \\
 & +\frac{U}{2}\sum_{k\sigma}[A_{k\sigma}^{\dagger}A_{k\sigma}+B_{k\sigma}^{\dagger}B_{k\sigma}-\sigma(A_{k\sigma}^{\dagger}B_{k\sigma}+B_{k\sigma}^{\dagger}A_{k\sigma})],
\end{align}
where\begin{equation}
\varepsilon_{k}=-2\sqrt{2}t\cos(k/2),
\end{equation}
with $k=2\pi j(\frac{3}{N})-\pi$, and $j=1,\ldots,N/3$. We can now exactly diagonalize $\mathcal{H}_{0}$ through the
following Bogoliubov transformation:
\begin{equation}
A_{k\sigma}=u_{k}\alpha_{k\sigma}-\sigma v_{k}\beta_{k\sigma},\,\,\,\,\,B_{k\sigma}=\sigma v_{k}\alpha_{k\sigma}+u_{k}\beta_{k\sigma},\label{Bogoliubov transformation}
\end{equation}
with $u_{k}$ and $v_{k}$ satisfying the canonical constraint: $(u_{k})^{2}+(v_{k})^{2}=1$, to maintain the  anticommutation relations of the Grassmann fields. Due to the
ferrimagnetic order of the GS, the above transformation is subject to a $4\pi$ periodicity of the Bogoliubov functions $\left\{ u_{k},v_{k}\right\} $
and Grassmann fields $\left\{ \alpha_{k\sigma},\beta_{k\sigma}\right\} $. The diagonalized
$\mathcal{H}_{0}$ thus reads: 
\begin{eqnarray}
\mathcal{H}_{0} & = & -\sum_{k\sigma}(E_{k}-\frac{U}{2})\alpha_{k\sigma}^{\dagger}\alpha_{k\sigma}+\sum_{k\sigma}(E_{k}+\frac{U}{2})\beta_{k\sigma}^{\dagger}\beta_{k\sigma}\nonumber \\
 &  & +\frac{U}{2}\sum_{k\sigma}(1-\sigma)e_{k\sigma}^{\dagger}e_{k\sigma},\label{eq:Diagonal}
\end{eqnarray}
where
\begin{equation}
(u_{k},v_{k})=\frac{1}{\sqrt{2}}\left(1\pm\frac{\left|\varepsilon_{k}\right|}{E_{k}}\right)^{1/2},\label{eq:uv}
\end{equation}
and
\begin{equation}
E_{k}=\sqrt{\varepsilon_{k}^{2}+U^{2}/4}.
\end{equation}
As one can see from Eq.~(\ref{eq:Diagonal}), the non-interacting
tight binding ($U=0$) spectrum of $\mathcal{H}_{0}$ present three
electronic bands: a nondispersive flat band (related to the Grassmann
fields $\{e_{k\sigma}^{\dagger},e_{k\sigma}\}$, macroscopically degenerate),
and two dispersive ones. In $AB_{2}$ chains, flat bands are closely
associated with ferrimagnetism (unsaturated ferromagnetism)~\cite{Macedo1995,Tian1996,MontenegroFilho2005}  at half filling, in agreement with Lieb's theorem~\cite{Lieb1989,Lieb1995hubbard}, or fully polarized ferromagnetism~\cite{Tasaki1998nagaoka} associated with the flat lowest band. We also stress that even at this level of approximation and in the weak coupling regime ($U=2t$), it was shown~\cite{Macedo1995} that hole doping [parametrized by $\delta$ defined in Eq.~(\ref{e23})] can destroy the ferrimagnetic order and/or induce phase separation in $AB_2$ chains. As depicted in Fig.~\ref{Fig_1}\subref{Fig_1a}, the $U=0$ spin degeneracy of the flat bands is removed by the Coulombian
repulsive interaction, in which case a gap $U$ opens between the
$e_{k\sigma}$ modes: $e_{k\uparrow}=0$, where spins at sites $B_{1}$
and $B_{2}$ are up, and $e_{k\downarrow}=U$, where these spins are down. On the other hand, the two dispersive bands are spin
degenerated, and also display a Hubbard gap $U$ separating the low $(\alpha_{k\sigma})$-energy
and high $(\beta_{k\sigma})$-energy modes~\cite{Oliveira_2009}.

\subsection{\label{Strong} Strong-coupling limit}

In this subsection, we shall introduce a perturbative expansion in
the strong-coupling regime ($U\gg t$) in order to obtain a low-energy
effective Lagrangian for the $AB_{2}$ Hubbard chain at half filling 
and in the doped regime. First, we resume the results of the previous section by writing the Grassmann fields $d_{i\sigma}$ and $b_{i\sigma}$ in terms of the Grassmann (Bogoliubov) fields $\alpha_{k\sigma}$ and $\beta_{k\sigma}$:
\begin{eqnarray}
(d_{i\sigma},b_{i\sigma}) & = & \frac{1}{\sqrt{2N_{c}}}\sum_{k}(e^{ikx_{i}},e^{ik(x_{i}-\frac{1}{2})})\nonumber \\
 &  & \times[(u_{k}\pm\sigma v_{k})\alpha_{k\sigma}\pm(u_{k}\mp\sigma v_{k})\beta_{k\sigma}],\label{BogoliubovField_di}
\end{eqnarray}where the phase
factor $e^{-\frac{ik}{2}}$ signalizes the quasi-1D $AB_{2}$ structure, and the antisymmetric Grassmann field $e_{i,\sigma}$ remains as defined in Eq.~(\ref{SymmetryProperty}). In the strong-coupling limit, however, it will prove useful to define a set of auxiliary spinless Grassmann fields~\cite{Weng_1991prl,Weng_1992prb,Oliveira_2009} in direct
space associated  with $d_{i\sigma}$ and $b_{i\sigma}$:
\begin{equation}
(\alpha_{i},\beta_{i})=  \sqrt{\frac{1}{N_{c}}}\sum_{k,\sigma}\theta(\pm\sigma)e^{ikx_{i}}(\alpha_{k\sigma},\beta_{k\sigma}),\label{GrassmannFieldsDirectSpace}\end{equation}and a similar equation for ${(\alpha_{i}^{\frac{1}{2}},\beta_{i}^{\frac{1}{2}})}\leftrightarrow(\alpha_{k\sigma},\beta_{k\sigma})$ is obtained by the replacements: $\theta(\pm\sigma)\rightarrow \theta(\mp\sigma)$ and $x_{i}\rightarrow x_{i}-1/2$, where $\theta(\sigma)$ is the Heaviside function, while for the antisymmetric component, one has
\begin{equation}
e_{i,\sigma}=\sqrt{\frac{1}{N_{c}}}\sum_{k}e^{ikx_{i}}e_{k,\sigma}.
\end{equation}
Now, by expanding $(u_{k},v_{k})$ in Eq.~(\ref{eq:uv}) in powers of $t/U$:
\begin{equation}
(u_{k},v_{k})\approx\frac{1}{\sqrt{2}}\left[1\pm\frac{\left|\varepsilon_{k}\right|}{U}+\mathcal{O}\left(\frac{t^{2}}{U^{2}}\right)\right],\label{eq:expansion}
\end{equation}
substituting these results into the Eq.~(\ref{BogoliubovField_di}), and using the inverse transformation
of Eq.~(\ref{GrassmannFieldsDirectSpace}), we can derive a perturbative
expansion in powers of $t/U$ for the Grassmann fields $d_{i\sigma}$
and $b_{i\sigma}$ in terms of the spinless Grassmann fields as follows:
\begin{align}
d_{i\sigma} & =\theta(\sigma)\alpha_{i}+\theta(-\sigma)\beta_{i}+\sqrt{2}\frac{t}{U}\theta(-\sigma)(\alpha_{i}^{\frac{1}{2}}+\alpha_{i+1}^{\frac{1}{2}})\nonumber \\
 & +\frac{t}{U}\theta(\sigma)[\sqrt{2}(\beta_{i}^{\frac{1}{2}}+\beta_{i+1}^{\frac{1}{2}})-\frac{t}{U}(2\alpha_{i}+\alpha_{i+1}+\alpha_{i-1})]\nonumber \\
 & +\mathcal{O}(t^{2}/U^{2}),\label{perturbative expansion_di}
\end{align}
\begin{align}
b_{i\sigma} & =\theta(-\sigma)\alpha_{i}^{\frac{1}{2}}-\theta(\sigma)\beta_{i}^{\frac{1}{2}}+\sqrt{2}\frac{t}{U}\theta(\sigma)(\alpha_{i}+\alpha_{i-1})\nonumber \\
 & -\frac{t}{U}\theta(-\sigma)[\sqrt{2}(\beta_{i}+\beta_{i-1})+\frac{t}{U}(2\alpha_{i}^{\frac{1}{2}}+\alpha_{i+1}^{\frac{1}{2}}+\alpha_{i-1}^{\frac{1}{2}})]\nonumber \\
 & +\mathcal{O}(t^{2}/U^{2}).\label{perturbative expansion_bi}
\end{align}
In the above derivation, we have used that $\theta(\sigma)\theta(\sigma')=\theta(\sigma)\delta_{\sigma,\sigma'}$.
Notice that, since  $\frac{t}{U}\ll1$, in Eqs.~(\ref{perturbative expansion_di})
and (\ref{perturbative expansion_bi}) we can identify the fields $\alpha_{i}^{\frac{1}{2}}\approx a_{iA\downarrow}$
and $\alpha_{i}\approx(a_{iB_{1}\uparrow}+a_{iB_{2}\uparrow})/\sqrt{2}$,
a result fully consistent with the low-energy spin configuration
of the ferrimagnetic state discussed previously. Analogously, for
the high-energy bands, the opposite spin configuration is observed,
with spin up (down) present at sites $A$ ($B_{1}$, $B_{2}$). 

Introducing Eqs.~(\ref{perturbative expansion_di}) and (\ref{perturbative expansion_bi})
into Eq.~(\ref{eq:Hamil01}), with the aid
 of Eq.~(\ref{SymmetryProperty}), we obtain a perturbative expression
for $\mathcal{H}_{0}$ (low-energy sector) in terms of the spinless Grassmann fields up to order $J=4t^{2}/U$:
\begin{multline}
\mathcal{H}_{0}=-J\sum_{i}[\alpha_{i}^{\dagger}\alpha_{i}+\alpha_{i}^{(\frac{1}{2})\dagger}\alpha_{i}^{\frac{1}{2}}-\beta_{i}^{\dagger}\beta_{i}-\beta_{i}^{(\frac{1}{2})\dagger}\beta_{i}^{\frac{1}{2}}]\\
-\frac{J}{2}\sum_{i}[\alpha_{i}^{\dagger}\alpha_{i+1}+\alpha_{i}^{(\frac{1}{2})\dagger}\alpha_{i+1}^{\frac{1}{2}}-\beta_{i}^{\dagger}\beta_{i+1}-\beta_{i}^{(\frac{1}{2})\dagger}\beta_{i+1}^{\frac{1}{2}}+\textrm{H.c}.]\\
+U\sum_{i}[\beta_{i}^{\dagger}\beta_{i}+\beta_{i}^{(\frac{1}{2})\dagger}\beta_{i}^{\frac{1}{2}}+e_{i\downarrow}^{\dagger}e_{i\downarrow}].\label{eq:H0}
\end{multline}
By applying Fourier transform to the above expression and rearranging the terms, we obtain\begin{eqnarray}
\mathcal{H}_{0} & = & -\sum_{k}2J\cos^{2}(k/2)(\alpha_{k}^{\dagger}\alpha_{k}+\alpha_{k}^{(\frac{1}{2})\dagger}\alpha_{k}^{\frac{1}{2}})\nonumber \\
 &  & +\sum_{k}[2J\cos^{2}(k/2)+U](\beta_{k}^{\dagger}\beta_{k}+\beta_{k}^{(\frac{1}{2})\dagger}\beta_{k}^{\frac{1}{2}})\nonumber \\
 &  & +\frac{U}{2}\sum_{k\sigma}(1-\sigma)e_{k\sigma}^{\dagger}e_{k\sigma}.\label{eq:H0-1}
\end{eqnarray}\begin{figure}
\centering
\subfloat{\label{Fig_1a}\includegraphics[scale=1]{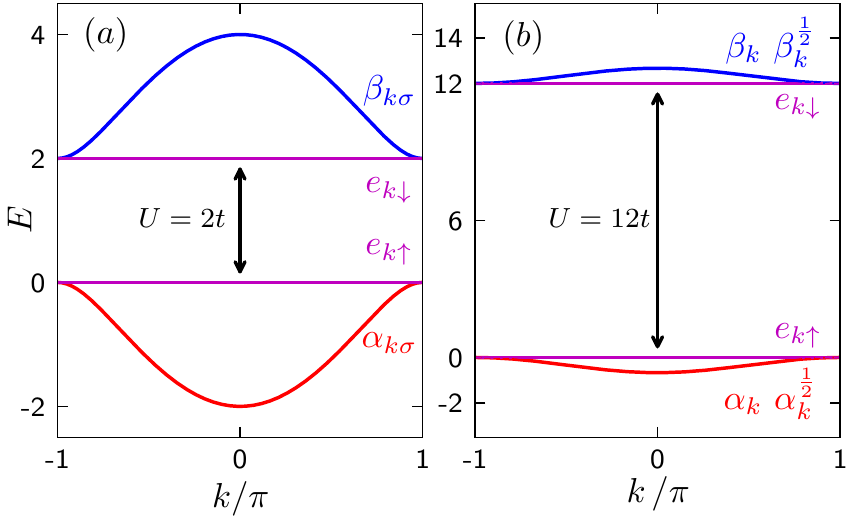}}
\subfloat{\label{Fig_1b}}
\caption{(Color online) Electronic spectrum of the Hamiltonian $\mathcal{H}_{0}$: (a) Eq.~(\ref{eq:Diagonal}) for $U=2t$  and (b)  Eqs.~(\ref{eq:Diagonal}) and (\ref{eq:H0-1}) for $U=12t$ ($J=4t^2/U=1/3$), with $t\equiv1$. Notice the band shrinking phenomenon as $U$ increases from $2t$ to $12t$ (strong-coupling regime). The $t\ll U$ expansion of the fields identifies $\alpha_{k}^{\frac{1}{2}}\approx a_{kA\downarrow}$,
$\alpha_{k}\approx(a_{kB_{1}\uparrow}+a_{kB_{2}\uparrow})/\sqrt{2}$
and $e_{k\uparrow}\approx(a_{kB_{1}\uparrow}-a_{kB_{2}\uparrow})/\sqrt{2}$,  where spins at sites $A$ ($B_{1},B_{2}$) 
are down (up), in  agreement with Lieb’s theorem~\cite{Lieb1989}.}
\label{Fig_1}
\end{figure}In Fig.~\ref{Fig_1} we plot the  electronic spectrum of the Hamiltonian $\mathcal{H}_{0}$, both in the weak and strong-coupling regime: (a) Eq.~(\ref{eq:Diagonal}) for $U=2t$  and (b)  Eqs.~(\ref{eq:Diagonal}) and (\ref{eq:H0-1}) for $U=12t$ ($J=4t^2/U=1/3$), respectively, with $t\equiv1$. We can notice the presence of the shrinking phenomenon~\cite{Macedo1995} as $U$ increases from $2t$ to $12t$ (strong-coupling regime) and that, for $U=12t$, Eq.~(\ref{eq:H0-1}) is a very good approximation to Eq.~(\ref{eq:Diagonal}). Noticeably, the $t\ll U$ expansion of the fields allow us to identify $\alpha_{k}^{\frac{1}{2}}\approx a_{kA\downarrow}$,
$\alpha_{k}\approx(a_{kB_{1}\uparrow}+a_{kB_{2}\uparrow})/\sqrt{2}$ (triplet state)
and $e_{k\uparrow}\approx(a_{kB_{1}\uparrow}-a_{kB_{2}\uparrow})/\sqrt{2}$ (singlet state), as the low-energy spin configuration of the ferrimagnetic state with single occupancy, where spins at sites $A$($B_{1},B_{2}$) 
are down (up), in  agreement with Lieb’s theorem~\cite{Lieb1989,Macedo1995,Tian1996}.

In order to describe the most relevant low-energy processes that take
place in this regime, one has to additionally project out the high-energy
bands from $\mathcal{H}_{0}$, that is, terms containing only fields
related to the high-energy bands are excluded. Therefore, after the Legendre transformation, $\mathcal{H}_{0}=-\sum_{i,\eta_{i}}\frac{\partial\mathcal{L}_{0}}{\partial(\partial_{\tau}\eta_{i})}\partial_{\tau}\eta_{i}+\mathcal{L}_{0}$,
where $\eta_{i}=\alpha_{i},\alpha_{i}^{\frac{1}{2}}$, and $e_{i\uparrow}$
(fields related to the low-energy bands), with $\frac{\partial\mathcal{L}_{0}}{\partial(\partial_{\tau}\eta_{i})}=\eta_{i}^{\dagger}$,
the Lagrangian associated with $\mathcal{H}_{0}$ (up to order $J$) is given by
\begin{eqnarray}
\mathcal{L}_{0} & = & \sum_{i,\eta_{i}}\eta_{i}^{\dagger}\partial_{\tau}\eta_{i}-J\sum_{i}(\alpha_{i}^{\dagger}\alpha_{i}+\alpha_{i}^{(\frac{1}{2})\dagger}\alpha_{i}^{\frac{1}{2}})\nonumber \\
 &  & -\frac{J}{2}\sum_{i}(\alpha_{i}^{\dagger}\alpha_{i+1}+\alpha_{i}^{(\frac{1}{2})\dagger}\alpha_{i+1}^{\frac{1}{2}}+\textrm{H.c}.).\label{eq:L0}
\end{eqnarray}

We shall now focus on the $U\gg t$ perturbative expansion
of $\mathcal{L}_{n}$, Eq.~(\ref{eq:LanN}), which amounts to consider
the most significant low-energy processes, after the use of Eqs.~(\ref{perturbative expansion_di})
and (\ref{perturbative expansion_bi}) for $d_{i\sigma}$
and $b_{i\sigma}$ in terms of the spinless Grassmann fields. However, terms allowing interband transitions
between low- and high-energy bands do exist in $\mathcal{L}_{n}$.  In this context, we apply a suitable second-order Rayleigh-Schr\"{o}dinger
perturbation theory~\cite{Weng_1991prl,Oliveira_2009}, consistent with the strong-coupling expansion, so that the modes associated with the high-energy bands are eliminated. Lastly, by adding $\mathcal{L}_{0}$ to the perturbative expansion of $\mathcal{L}_{n}$, which leads to the cancellation  of the exchange terms in Eq.~(\ref{eq:L0}), the effective low-energy Lagrangian density of the $AB_{2}$ Hubbard model in the strong-coupling limit (up to order $J$) reads:

\begin{equation}
\mathcal{L}_{eff}(\tau)=\mathcal{L}^{(I)}+\mathcal{L}^{(II)}+\mathcal{L}^{(III)}+\mathcal{L}^{(IV)},\label{eq:Leffective}
\end{equation}
where\begin{subequations} \label{eq:Lagrangianwhole}
\begin{equation}
\mathcal{L}^{(I)}=\sum_{i}\alpha_{i}^{\dagger}\partial_{\tau}\alpha_{i}+\sum_{i}\alpha_{i}^{(\frac{1}{2})\dagger}\partial_{\tau}\alpha_{i}^{(\frac{1}{2})}+\sum_{i}e_{i\uparrow}^{\dagger}\partial_{\tau}e_{i\uparrow},\label{32a}
\end{equation}
\begin{align}
\mathcal{L}^{(II)}  =&\sum_{i\sigma}\left\{ \theta(-\sigma)(U_{i}^{(b)\dagger}\partial_{\tau}U_{i}^{(b)})_{\sigma,\sigma}\alpha_{i}^{(\frac{1}{2})\dagger}\alpha_{i}^{(\frac{1}{2})}\right.\nonumber \\
 & +\theta(\sigma)\frac{1}{2}[(U_{i}^{(d)\dagger}\partial_{\tau}U_{i}^{(d)})_{\sigma,\sigma}+(U_{i}^{(e)\dagger}\partial_{\tau}U_{i}^{(e)})_{\sigma,\sigma}]\nonumber \\
 & \times(\alpha_{i}^{\dagger}\alpha_{i}+e_{i\uparrow}^{\dagger}e_{i\uparrow})+\left[\theta(\sigma)\frac{1}{2}[(U_{i}^{(d)\dagger}\partial_{\tau}U_{i}^{(e)})_{\sigma,\sigma}\right.\nonumber \\
 & \left.\left.+(U_{i}^{(e)\dagger}\partial_{\tau}U_{i}^{(d)})_{\sigma,\sigma}]\alpha_{i}^{\dagger}e_{i\uparrow}+\textrm{H.c.}\right]\right\} ,\label{32b}
\end{align}
\begin{align}
\mathcal{L}^{(III)}= & -t\sum_{i\sigma}\left\{ \theta(-\sigma)(U_{i}^{(b)\dagger}U_{i}^{(d)})_{\sigma,-\sigma}\alpha_{i}^{(\frac{1}{2})\dagger}\alpha_{i}\right.\nonumber \\
 & +\theta(\sigma)(U_{i}^{(d)\dagger}U_{i+1}^{(b)})_{\sigma,-\sigma}\alpha_{i}^{\dagger}\alpha_{i+1}^{(\frac{1}{2})}\nonumber \\
 & +\theta(-\sigma)(U_{i}^{(b)\dagger}U_{i}^{(e)})_{\sigma,-\sigma}\alpha_{i}^{(\frac{1}{2})\dagger}e_{i\uparrow}\nonumber \\
 & \left.+\theta(\sigma)\left(U_{i}^{(e)\dagger}U_{i+1}^{(b)}\right)_{\sigma,-\sigma}e_{i\uparrow}^{\dagger}\alpha_{i+1}^{(\frac{1}{2})}+\textrm{H.c.}\right\} ,\label{32c}
\end{align}
\begin{align}
\mathcal{L}^{(IV)}= & -\frac{J}{4}\sum_{i;i'=i,i+1;\sigma}\theta(\sigma)|(U_{i}^{(d)\dagger}U_{i'}^{(b)})_{\sigma,\sigma}|^{2}\alpha_{i}^{\dagger}\alpha_{i}\nonumber \\
 & -\frac{J}{4}\sum_{i;i'=i,i+1;\sigma}\theta(\sigma)|(U_{i}^{(e)\dagger}U_{i'}^{(b)})_{\sigma,\sigma}|^{2}e_{i\uparrow}^{\dagger}e_{i\uparrow}\nonumber \\
 & -\frac{J}{4}\sum_{i;i'=i,i-1;\sigma}\theta(-\sigma)[|(U_{i}^{(b)\dagger}U_{i'}^{(d)})_{\sigma,\sigma}|^{2}\nonumber \\
 & +(U_{i}^{(b)\dagger}U_{i'}^{(e)})_{\sigma,\sigma}|^{2}]\alpha_{i}^{(\frac{1}{2})\dagger}\alpha_{i}^{(\frac{1}{2})},\label{32d}
\end{align}
\end{subequations}where
\begin{equation}
U_{i}^{(b)}=U_{iA},\,\,\,
U_{i}^{(d,e)}=\frac{1}{\sqrt{2}}(U_{iB_{1}}\pm U_{iB_{2}}),\label{SymmetryPropertyanother}
\end{equation}
in which case we took advantage of the symmetry of
the $AB_{2}$ chain under the exchange operation $B_{1}\leftrightarrow B_{2}$,
in correspondence with Eq.~(\ref{SymmetryProperty}). From the above equations, we see that the kinetic
term is represented by $\mathcal{L}^{(I)}$ and is related to the
charge degrees of freedom only, whereas $\mathcal{L}^{(II)}$ describes
the dynamics of the spin degrees of freedom coupled to the charge
fields. On the other hand, $\mathcal{L}^{(III)}$ exhibit  first-neighbor hopping contributions between charge degrees of freedom in the presence of $SU(2)$ gauge fields, while $\mathcal{L}^{(IV)}$  is the spin exchange term in the  presence of the charge Grassmann  fields.

\section{\label{sec:HALF-FILLING} half-filling regime}

Let us now discuss some basic aspects of the localized magnetic properties
related to the spin degrees of freedom. At half filling, i.e., $\delta=0$,
 we have $\langle\alpha_{i}^{\dagger}\alpha_{i}\rangle=1$,
$\langle\alpha_{i}^{\left(1/2\right)\dagger}\alpha_{i}^{(1/2)}\rangle=1$,
$\langle e_{i\uparrow}^{\dagger}e_{i\uparrow}\rangle=1$, and $\langle\alpha_{i}^{\dagger}e_{i\uparrow}\rangle=0$
(no band hybridization) as the electrons tend to fill up the lower-energy
bands, whereas the higher-energy ones remain empty. As a consequence,
a  ferrimagnetic configuration of localized spins emerges,
i.e., the charge degrees of freedom are completely frozen, such that
$\langle\alpha_{i}^{\dagger}\partial_{\tau}\alpha_{i}\rangle=\langle\alpha_{i}^{\left(1/2\right)\dagger}\partial_{\tau}\alpha_{i}^{(1/2)}\rangle=\langle e_{i\uparrow}^{\dagger}\partial_{\tau}e_{i\uparrow}\rangle=0$,
with forbidden hopping. Therefore, only terms from $\mathcal{L}^{II}$
and $\mathcal{L}^{IV}$ in Eqs.~(\ref{32b}) and (\ref{32d}), respectively, give nonzero contributions and the resulting effective strong-coupling Lagrangian at half filling, defined in Eq.~(\ref{eq:Leffective}),  reads:
\begin{eqnarray}
 & \mathcal{L}_{eff}^{J}= & \sum_{i\alpha\sigma}\theta(p_{i\alpha}\sigma)(U_{i\alpha}^{\dagger}\partial_{\tau}U_{i\alpha})_{\sigma,\sigma}\nonumber \\
 &  & -\frac{J}{4}\sum_{\left\langle i\alpha,j\beta\right\rangle \sigma}\theta(p_{i\alpha}\sigma)\left|(U_{i\alpha}^{\dagger}U_{j\beta})_{\sigma,\sigma}\right|^{2},\label{Leffheisenberg}
\end{eqnarray}
where the staggered factor $p_{i\alpha}$ was defined in Eq.~(\ref{eq4}), and use was made of the matrix transformations defined  in Eq.~(\ref{SymmetryPropertyanother})  in order to sum up the squares  of the $SU(2)$ gauge field products in the exchange contribution
from $\mathcal{L}^{IV}$ in Eq.~(\ref{32d}). Now, using the following Legendre transform: $\mathcal{H}_{eff}^{J}=-\sum_{i\alpha\sigma}\frac{\partial\mathcal{L}_{eff}^{J}}{\partial(\partial_{\tau}U_{i\alpha})_{\sigma,\sigma}}(\partial_{\tau}U_{i\alpha})_{\sigma,\sigma}+\mathcal{L}_{eff}^{J},\label{LegendreHeisen}$ where $\frac{\partial\mathcal{L}_{eff}^{J}}{\partial(\partial_{\tau}U_{i\alpha})_{\sigma,\sigma}}=\theta(p_{i\alpha}\sigma)(U_{i\alpha}^{\dagger})_{\sigma,\sigma},$
we get the respective quantum Heisenberg Hamiltonian written in terms
of the $SU(2)$ gauge fields at half filling as
\begin{equation}
\mathcal{H}_{eff}^{J}=-\frac{J}{4}\sum_{\left\langle i\alpha,j\beta\right\rangle \sigma}\theta(p_{i\alpha}\sigma)\left|(U_{i\alpha}^{\dagger}U_{j\beta})_{\sigma,\sigma}\right|^{2}.\label{eq:Heisenberglike}
\end{equation}
Further, using the definition of the $SU(2)/U(1)$ unitary rotation
matrix Eq.~(\ref{Ufor}), it is possible to write~\cite{Weng_1991prl,Weng_1992prb,Oliveira_2009,Ribeiro2015}
$\left|(U_{i\alpha}^{\dagger}U_{j\beta})_{\sigma,\sigma}\right|^{2}=\frac{1}{2}(1+\mathbf{n}_{i\alpha}\cdot\mathbf{n}_{j\beta}),$
where $\mathbf{n}_{i\alpha}=\sin(\theta_{i\alpha})\left[\cos(\phi_{i\alpha})\hat{\mathbf{x}}+\sin(\phi_{i\alpha})\hat{\mathbf{y}}\right]+\cos(\theta_{i\alpha})\hat{\mathbf{z}}$
is the unit vector pointing along the local spin direction. Lastly, by using the constraint as given in Eq.~(\ref{stilda}), we can identify the spin field $\{\mathbf{S}_{i\alpha}\}$ at the single occupied sites: \begin{equation}
\mathbf{S}_{i\alpha}=p_{i\alpha}\mathbf{n}_{i\alpha}/2,\label{Spialpha}
\end{equation}where $p_{i\alpha}=+1\,\,(-1)$ at sites $\alpha=B_{1},\,B_{2}$ $(A)$, in order to obtain\begin{align}
\mathcal{H}_{eff}^{J} & =J\sum_{i}\left[(\mathbf{S}_{i}^{B_{1}}+\mathbf{S}_{i}^{B_{2}})\cdot(\mathbf{S}_{i}^{A}+\mathbf{S}_{i+1}^{A})\right]-JN_{c}.\label{eq:heisenbergeff}
\end{align}
The above expression is indeed that of the quantum antiferromagnetic Heisenberg spin-1/2 model on the $AB_{2}$ chain in zero-field, which takes into account the effects
of zero-point quantum spin fluctuations. In fact, to achieve this goal, we analyze the Hamiltonian,
Eq.~(\ref{eq:heisenbergeff}), by means of the spin-wave theory,
which has proved very successful in describing the properties of the
GS and low-lying excited states of spin models.  The predicted results provide a check of the consistency of our approach and will be fully used in our description of the doped regime.

We shall first introduce boson creation and annihilation operators via the Holstein-Primakoff~\cite{fradkin2013field} transformation:\begin{equation}
\begin{aligned}
   S_{i}^{A,z}=&-S+a_{i}^{\dagger}a_{i}, \\     
   S_{i}^{A,+}=&\,(S_{i}^{A,-})^{\dagger}=\sqrt{2S}a_{i}^{\dagger}f_{A}(S),\label{eq:Holstein-PrimakoffA}
\end{aligned}
\end{equation}
for a down-spin on the $A$ site, and
\begin{equation}
\begin{aligned}
   S_{i}^{B_{l},z}&=S-b_{li}^{\dagger}b_{li},\\   
   S_{i}^{B_{l},+}&=(S_{i}^{B_{l},-})^{\dagger}=\sqrt{2S}f_{B}(S)b_{li},\label{eq:Holstein-PrimakoffB}
\end{aligned}
\end{equation}
for an up-spin on the $B_{l}$ site, with $l=1,2$, and 
\begin{equation}
f_{r}(S)=\left(1-\frac{n_{r}}{2S}\right)^{1/2}=1-\frac{1}{2}\frac{n_{r}}{2S}+\ldots,\label{expansion}
\end{equation}
where $S$ is the spin magnitude, and $n_{r}=a_{i}^{\dagger}a_{i}$
or $b_{li}^{\dagger}b_{li}$. The operators $a_{i}^{\dagger}$ and
$a_{i}$ (or $b_{li}^{\dagger}$, $b_{li}$ ) satisfy the boson commutation
rules. Under the
above transformation, the spin Hamiltonian, Eq.~(\ref{eq:heisenbergeff})
is mapped onto the boson Hamiltonian:
\begin{equation}
\mathcal{H}_{eff}^{J}=E_{0}-JN_{c}+\mathcal{H}_{1}+\mathcal{H}_{2}+\mathcal{O}(S^{-1}),\label{eq:BosonHamiltonian}
\end{equation}
where 
\begin{equation}
E_{0}=-4S^{2}JN_{c},\label{eq:E0}
\end{equation}
is the classical GS energy and $\mathcal{H}_{1}$ and $\mathcal{H}_{2}$
are the quadratic and quartic (interacting)
terms of the boson Hamiltonian, suitable
to describe the quantum $AB_{2}$ Heisenberg model 
via a perturbative series expansion in powers of $1/S$.  By Fourier transforming the boson operators, we find\begin{eqnarray}
\mathcal{H}_{1}= &  & \,2JS\sum_{k}(2a_{k}^{\dagger}a_{k}+\sum_{l}b_{lk}^{\dagger}b_{lk})\nonumber \\
 &  & +\sum_{k,l=1,2}2JS\gamma_{k}(a_{k}^{\dagger}b_{lk}^{\dagger}+a_{k}b_{lk}),
\end{eqnarray}
where we have defined the lattice structure factor as 
\begin{equation}
\gamma_{k}=\frac{1}{z}\sum_{\rho}e^{ik\rho}=\cos(\frac{k}{2}),
\end{equation}
with $z$ denoting the coordination number ($z=4$ for the $AB_{2}$
chain), while $\rho=\pm1/2$ connects the nearest neighbors $A\textrm{-}B_{1}$
and $A\textrm{-}B_{2}$ linked sites of sublattices $A$ and $B$,
and
\begin{align}
\mathcal{H}_{2} & =-\frac{3J}{2N}\sum_{1234,l=1,2}\delta_{12,34}\left\{ 4\gamma_{1-4}a_{1}^{\dagger}a_{4}b_{l3}^{\dagger}b_{l2}\right.\nonumber \\
 & \left.+(\gamma_{1}a_{1}^{\dagger}b_{l4}^{\dagger}b_{l3}^{\dagger}b_{l2}+\gamma_{1+2-3}a_{1}^{\dagger}a_{2}^{\dagger}a_{3}b_{l4}^{\dagger}+\textrm{H.c.})\right\}.\label{eq: quartic boson interacting Hamiltonian}
\end{align}
For simplicity, we
use the convention $1$ for $k_{1}$, $2$ for $k_{2}$, and so on.
Also, the $\delta_{12,34}=\delta(k_{1}+k_{2}-k_{3}-k_{4})$ is the
Kronecker $\delta$ function, and expresses the conservation of momentum
to within a reciprocal-lattice vector $G$. 

We shall consider $\mathcal{H}_{1}$ first, which is the term leading to linear spin-wave theory (LSWT). In fact, $\mathcal{H}_{1}$ is diagonalized using the following  Bogoliubov transformation:
\begin{equation}
\begin{aligned}
a_{k} & = u_{k}\beta_{k}-v_{k}\alpha_{k}^{\dagger}, \\  
b_{lk} & = \frac{1}{\sqrt{2}}[u_{k}\alpha_{k}-v_{k}\beta_{k}^{\dagger}+(-1)^l\xi_{k}],\text{ with $l=1,2$},\label{eq:Bogoliubov transformation}
\end{aligned}
\end{equation}
\begin{equation}
(u_{k},v_{k})=\frac{(3+\sqrt{9-8\gamma_{k}^{2}},\,\,\,2\sqrt{2}\gamma_{k})}{\sqrt{(3+\sqrt{9-8\gamma_{k}^{2}})^{2}-8\gamma_{k}^{2}}},\label{BogoliubovUVk}
\end{equation}
where $u_k$ and $v_k$ satisfy the constraint $u_{k}^{2}-v_{k}^{2}=1$. Thus, 
\begin{equation}
\mathcal{H}_{1}=E_{1}+\sum_{k}(\epsilon_{k}^{0(\alpha)}\alpha_{k}^{\dagger}\alpha_{k}+\epsilon_{k}^{0(\beta)}\beta_{k}^{\dagger}\beta_{k}+\epsilon_{k}^{0(\xi)}\xi_{k}^{\dagger}\xi_{k});\label{eq:BosonHamiltonian1}
\end{equation}
\begin{equation}
E_{1}=JS\sum_{k}(\sqrt{9-8\gamma_{k}^{2}}-3),\label{eq:E1}
\end{equation}
\begin{equation}
\epsilon_{k}^{0(\alpha,\beta)}=JS(\sqrt{9-8\gamma_{k}^{2}}\mp1),\,\,\,\,\,\,\,\epsilon_{k}^{0(\xi)}=2JS,\label{eq:espinwave}
\end{equation}
where $E_{1}$ is the $\mathcal{O}(S^{1})$ quantum correction to
the GS energy, and $\epsilon_{k}^{0(\alpha,\beta)}$, $\epsilon_{k}^{0(\xi)}$
are the three spin-wave branches provided by LSWT, both in agreement
with previous results~\cite{Vitoriano2002,Yamamoto2007}.
In fact, it is well known that systems with a ferrimagnetic GS naturally
have ferromagnetic and antiferromagnetic spin-wave modes as their
elementary magnetic excitations (magnons).
For the $AB_{2}$ chain, there are three spin-wave branches: an antiferromagnetic
mode ($\epsilon_{k}^{0(\beta)}$) and two ferromagnetic ones ($\epsilon_{k}^{0(\alpha)}$
and $\epsilon_{k}^{0(\xi)}$). The mode $\epsilon_{k}^{0(\alpha)}$
is gapless at $k=0$, i.e., the Goldstone mode, with a quadratic (ferromagnetic)
dispersion relation $\epsilon_{k}^{0(\alpha)}\sim k^{2}$. The other
two modes are gapped. Notice that the gapped ferromagnetic mode $\epsilon_{k}^{0(\xi)}$
is flat, and is closely associated with ferrimagnetic properties at
half filling~\cite{Macedo1995,Tasaki1998nagaoka}. Since the dispersive
modes preserve the local triplet bond, they are identical to those
found in the spin-$1/2$ - spin-1 chains~\cite{PatiGSE,Yamamoto1998,Ivanov2000,Yamamoto2004}. These chains also exhibit interesting field-induced Luttinger liquid behavior~\cite{welinton_2017}.

Now, our aim is to obtain the leading corrections to LSWT, i.e., second-order spin-wave theory to the GS energy, sublattice magnetizations and Lieb GS total spin per unit cell. In doing so, we develop a perturbative
scheme for the description of this quartic term. First, we decompose
the two-body terms by means of the Wick theorem, via normal-ordering
protocol for boson operators. Conservation of momentum to within a
reciprocal-lattice vector, implies: $k_{1}=k+q$, $k_{2}=p-q$, $k_{3}=k$
and $k_{4}=p$. Then, we need to look at the
possible pairings of the 4 operators, as for example, in the first term
of Eq.~(\ref{eq: quartic boson interacting Hamiltonian}):
\global\long\def\bC#1#2{\begC#2{#1}}
\global\long\def\mC#1{\conC{#1}}
\global\long\def\eC#1#2{\endC#2{#1}}
\[
\bC{a_{k+q}^{\dagger}}1\eC{a_{p}}1\bC{b_{l,k}^{\dagger}}2\eC{b_{l,p-q}}2,\,\,\,\,\bC{a_{k+q}^{\dagger}}1\bC{a_{p}}2\eC{b_{l,k}^{\dagger}}1\eC{b_{l,p-q}}2,\,\,\,\,\bC{a_{k+q}^{\dagger}}1\bC{a_{p}}2\eC{b_{l,k}^{\dagger}}2\eC{b_{l,p-q}}1.
\]
Under this procedure, and by substituting the Bogoliubov
transformation, Eqs.~(\ref{eq:Bogoliubov transformation})-(\ref{BogoliubovUVk}), into Eq.~(\ref{eq: quartic boson interacting Hamiltonian}), we find

\begin{equation}
\mathcal{H}_{2}=E_{2}+\sum_{k}(\delta\epsilon_{k}^{(\alpha)}\alpha_{k}^{\dagger}\alpha_{k}+\delta\epsilon_{k}^{(\beta)}\beta_{k}^{\dagger}\beta_{k}+\delta\epsilon_{k}^{(\xi)}\xi_{k}^{\dagger}\xi_{k}),\label{correction}
\end{equation}
where
\begin{equation}
E_{2}/N_{c}=-2J(q_{1}^{2}+q_{2}^{2}-\frac{3}{\sqrt{2}}q_{1}q_{2}),\label{E2}
\end{equation}
and the corresponding corrections for the spin-wave dispersion relations read:
\begin{align}
\delta\epsilon_{k}^{(\alpha)}=\,\, & J[u_{k}^{2}(\sqrt{2}q_{2}-2q_{1})+2v_{k}^{2}(\sqrt{2}q_{2}-q_{1})]\nonumber \\
 & +4J\gamma_{k}u_{k}v_{k}\left[\frac{3}{2\sqrt{2}}q_{1}-q_{2}\right]+\mathcal{O}(S^{-1}),\label{dispersioncorrection2}
\end{align}
$\delta\epsilon_{k}^{(\beta)}$ is obtained from $\delta\epsilon_{k}^{(\alpha)}$
through the exchange of $u_{k}\leftrightarrow v_{k}$, and 
\begin{align}
\delta\epsilon_{k}^{(\xi)} & =J(\sqrt{2}q_{2}-2q_{1})+\mathcal{O}(S^{-1}).\label{dispersioncorrection3}
\end{align}
In Eqs.~(\ref{E2})-(\ref{dispersioncorrection3}) above, the quantities $q_{1}$ and $q_{2}$ are defined
by (thermodynamic limit)
\begin{equation}
q_{1}=\frac{1}{2\pi}\int_{-\pi}^{\pi}dk(v_{k}^{2}),\,\,\,\,\,\,\,\,q_{2}=\frac{1}{2\pi}\int_{-\pi}^{\pi}dk(\gamma_{k}u_{k}v_{k}).\label{eq:q1q2}
\end{equation}
We remark that in deriving Eqs.~(\ref{E2})-(\ref{dispersioncorrection3}), we
have neglected terms  containing anomalous products, such as, $\alpha_{k}^{\dagger}\beta_{k}^{\dagger}$
and vertex corrections.

Lastly, the above results of our perturbative $1/S$ series
expansion lead to the effective Hamiltonian:
\begin{equation}
\mathcal{H}_{eff}^{J}=E_{GS}^{J}-JN_{c}+\sum_{k}(\epsilon_{k}^{\alpha}\alpha_{k}^{\dagger}\alpha_{k}+\epsilon_{k}^{(\beta)}\beta_{k}^{\dagger}\beta_{k}+\epsilon_{k}^{(\xi)}\xi_{k}^{\dagger}\xi_{k}),\label{eq:HJeff}
\end{equation}
where 
\begin{equation}
E_{GS}^{J}=E_{0}+E_{1}+E_{2},\label{eq:EJGS}
\end{equation}
which can be read from Eqs.~(\ref{eq:E0}), (\ref{eq:E1}), and (\ref{E2}), respectively,
is the second-order result up to $\mathcal{O}(1/S)$ for the GS energy, and 
\begin{equation}
\epsilon_{k}^{(s)}=\epsilon_{k}^{0(s)}+\delta\epsilon_{k}^{(s)},\,\,\,\,\,\mathrm{with\,\,\,\,\,}s=\alpha,\,\beta,\,\xi,\label{eq:SpinWaveDispersion}
\end{equation}
are the corresponding second-order spin-wave modes, where the linear and  the second-order correction terms are given by  Eq.~(\ref{eq:espinwave}) and Eqs.~(\ref{dispersioncorrection2})-(\ref{eq:q1q2}), respectively.

\subsection{Second-order spin-wave analysis}
Our perturbative $1/S$ series expansion approach
is able to improve the LSWT result for the gap $\Delta=J$ of the
antiferromagnetic mode, which
should be compared with the second-order result derived from $\epsilon_{k}^{(\beta)}$, Eqs.~(\ref{eq:espinwave}), (\ref{dispersioncorrection2}) and (\ref{eq:SpinWaveDispersion}), at $k=0$: $\Delta=(1+\sqrt{2}q_{2})J\simeq1.676J$, 
in full agreement with similar spin-wave calculations
for $AB_2$~\cite{Yamamoto2007} and spin-$1/2$-spin-$1$~\cite{Ivanov2000,Yamamoto2004} chains, and in agreement
with numerical estimates using exact diagonalization, $\Delta=1.759J$, for both  $AB_2$~\cite{MontenegroFilho2005} and spin-$1/2$-spin-$1$~\cite{Yamamoto1998} chains. On the other hand, the LSWT predicts a gap $\Delta_{flat}=J$
for the flat ferromagnetic mode ($\epsilon_{k}^{(\xi)}$) in $AB_2$ chain, whereas our second-order spin-wave theory finds, using Eqs.~(\ref{eq:espinwave}), (\ref{dispersioncorrection3}) and (\ref{eq:SpinWaveDispersion}): $\Delta_{flat}=(1-2q_{1}+\sqrt{2}q_{2})J\simeq1.066J$,
in full agreement with a similar spin-wave procedure~\cite{Yamamoto2007}.
Surprisingly, the estimated value from Exact Diagonalization (ED)~\cite{MontenegroFilho2005}:
$\Delta_{flat}=1.0004J$, lies between these two theoretical values.
In fact, analytical approaches are still unable to reproduce the observed level crossing found in numerical calculations~\cite{MontenegroFilho2005,Yamamoto2007}
for the two ferromagnetic modes. This is probably due to the fact
that the different symmetries exhibited by the localized excitation
(flat mode) and the ferromagnetic dispersive mode are not explicitly manifested in the analytical approaches, so the levels avoid the crossing.

\subsection{Ground state energy }

In the thermodynamic limit, the second-order result for the GS energy of the $AB_{2}$ chain
per unit cell reads:
\begin{eqnarray}
\frac{E_{GS}^{J}}{N_{c}} & = & -4JS^{2}+\frac{JS}{2\pi}\int_{-\pi}^{\pi}dk\left(\sqrt{9-8\gamma_{k}^{2}}-3\right)\nonumber \\
 &  & -2J(q_{1}^{2}+q_{2}^{2}-\frac{3}{\sqrt{2}}q_{1}q_{2}).\label{eq:EGSN}
\end{eqnarray}
We remark that, at half filling, we shall not consider
the  constant term $-JN_c$ in Eq.~(\ref{eq:BosonHamiltonian}), with the purpose of comparison with preceding results. Performing the integration over the first BZ and taking $S=1/2$, we obtain that
the GS energy per site at zero-field is given by $-0.4869J$. This
result agrees very well with values obtained  using exact diagonalization~\cite{TakanoGSE}
($-0.485J$) and DMRG~\cite{NiggemannGSE} ($-0.4847J$) techniques.
For the spin-$1/2$ - spin-1 chain, the value obtained using
DMRG~\cite{PatiGSE} is $-0.72704J$. To compare it with our finding,
we need to multiply this value by $2/3$ (ratio between the number of sites of the two chains), yielding $-0.48469J$. 

\subsection{\label{Sublattice-magnetizations}Sublattice magnetizations and Lieb GS total spin per unit cell}

In order to derive results beyond LSWT, we introduce staggered magnetic
fields coupled to spins $S_{i}^{A,z}$ and $S_{i}^{B_{l},z}$, with $l=1,2$, through
the Zeeman terms: $-h_{A}\sum_{i}S_{i}^{A,z}$ and $-h_{B_{l}}\sum_{i}S_{i}^{B_{l},z}$,
which are added to $\mathcal{H}_{eff}^{J}$ in Eq.~(\ref{eq:heisenbergeff}).
Thus, $\left\langle \right.S^{A,z}\left.\right\rangle$ and  $\left\langle \right.S^{B_{l},z}\left.\right\rangle$ corresponding to sublattices $A$ and $B_{l}$  are obtained from $\left\langle \right.S^{A,z}\left.\right\rangle=-(1/N_{c})\sum_{i=1,2}[\partial E_{i}(h_{A})/\partial h_{A}]|_{h_{A}=0}$, and an analogous equation for $\left\langle \right.S^{B_{l},z}\left.\right\rangle$ using Eqs.~(\ref{eq:E1}) and (\ref{E2}):
\begin{multline}
(\left\langle \right.S^{A,z}\left.\right\rangle,\left\langle \right.S^{B_{l},z}\left.\right\rangle)=\mp S\pm\left(\frac{1}{2},\frac{1}{4}\right)\frac{1}{\pi}\int_{-\pi}^{\pi}dkv_{k}^{2}\\
\mp\left(\frac{1}{2},\frac{1}{4}\right)\frac{q_{1}}{\pi S}\int_{-\pi}^{\pi}dk\frac{\gamma_{k}^{2}}{(9-8\gamma_{k}^{2})^{3/2}}+\mathcal{O}(\frac{1}{S^{2}}).\label{SA}
\end{multline}Carrying out the above integration, we obtain $\left\langle \right.S^{A,z}\left.\right\rangle=-0.316343$ and
$\left\langle \right.S^{B_{l},z}\left.\right\rangle=0.408172$. These results are in good agreement with those
obtained using DMRG~\cite{Sierra1999} and ED~\cite{MontenegroFilho2005}
techniques: $\left\langle \right.S^{A,z}\left.\right\rangle=-0.2925$ and $\left\langle \right.S^{B_{l},z}\left.\right\rangle=0.3962$,
respectively, and with values for $\left\langle \right.S^{A,z}\left.\right\rangle$ and $2\left\langle \right.S^{B_{l},z}\left.\right\rangle$
for the spin-$1/2$ - spin-1 chain~\cite{PatiGSE,Yamamoto1998,Ivanov2000,Yamamoto2004}.
Although at zero temperature, the sublattice magnetizations are strongly
reduced by quantum fluctuations, as compared with their classical values,
the unit cell magnetization remains $S_{L}\equiv1/2$, where $S_{L}$
is the Lieb GS total spin per unit cell, in full agreement with Lieb's
theorem~\cite{Lieb1989,MontenegroFilho2005} for bipartite lattices:
\begin{equation}
S_{L}=\frac{1}{2}\left\Vert N_{A}-N_{B}\right\Vert ,
\end{equation}
with $N_{A}(N_{B})$ denoting the total number of spins in sublattice
$A(B)$ per unit cell.

\section{\label{sec:t-J}$\mathbf{t\textrm{-}J}$ Hamiltonian: Doping-Induced Phases, Ground state 
Energy and Total Spin}

In this section, we shall derive the corresponding $t$-$J$ Hamiltonian suitable to describe the strongly correlated $AB_{2}$ Hubbard chain in the doped regime, in which case both charge (Grassmann fields) and spin [$SU(2)$ gauge fields] quantum fluctuations are considered
on an equal footing. Indeed, the $t$-$J$ Hamiltonian can be derived
by means of the following Legendre transformation to Eq.~(\ref{eq:Leffective}):
\begin{eqnarray}
\mathcal{H}_{eff}^{t\textrm{-}J} & = & -\sum_{i,\mu=b,d,e}\frac{\partial\mathcal{L}_{eff}}{\partial(\partial_{\tau}U_{i}^{(\mu)})_{\sigma,\sigma}}(\partial_{\tau}U_{i}^{(\mu)})_{\sigma,\sigma}\nonumber \\
 &  & -\sum_{i,\nu_{i}}\frac{\partial\mathcal{L}_{eff}}{\partial(\partial_{\tau}\nu_{i})}\partial_{\tau}\nu_{i}+\mathcal{L}_{eff},
\end{eqnarray}
where ${\frac{\partial\mathcal{L}_{eff}}{\partial(\partial_{\tau}\nu_{i})}=\nu_{i}^{\dagger}}$ with ${\nu_{i}=\alpha_{i},\alpha_{i}^{\frac{1}{2}},e_{i\uparrow}}$; $\frac{\partial\mathcal{L}_{eff}}{\partial(\partial_{\tau}U_{i}^{(b)})_{\sigma,\sigma}}=\theta(-\sigma)(U_{i}^{(b)\dagger})_{\sigma,\sigma}\alpha_{i}^{(\frac{1}{2})\dagger}\alpha_{i}^{(\frac{1}{2})}$, and $\frac{\partial\mathcal{L}_{eff}}{\partial(\partial_{\tau}U_{i}^{(d,e)})_{\sigma,\sigma}}  =  \theta(\sigma)\frac{1}{2}[(U_{i}^{(d,e)\dagger})_{\sigma,\sigma}(\alpha_{i}^{\dagger}\alpha_{i}+e_{i\uparrow}^{\dagger}e_{i\uparrow})+(U_{i}^{(e,d)\dagger})_{\sigma,\sigma}(\alpha_{i}^{\dagger}e_{i\uparrow}+e_{i\uparrow}^{\dagger}\alpha_{i})]$, from which we can write the effective $t$-$J$ Hamiltonian as

\begin{equation}
\mathcal{H}_{eff}^{t\textrm{-}J}=\mathcal{H}^{t}+\mathcal{H}^{J},\label{eq:Htjeff}
\end{equation}
where 
\begin{align}
\mathcal{H}^{t}= & -t\sum_{i\sigma}\{\theta(-\sigma)(U_{i}^{(b)\dagger}U_{i}^{(d)})_{\sigma,-\sigma}\alpha_{i}^{\left(1/2\right)\dagger}\alpha_{i}\nonumber \\
 & +\theta(\sigma)(U_{i}^{(d)\dagger}U_{i+1}^{(b)})_{\sigma,-\sigma}\alpha_{i}^{\dagger}\alpha_{i+1}^{(1/2)}\nonumber \\
 & +\theta(-\sigma)(U_{i}^{(b)\dagger}U_{i}^{(e)})_{\sigma,-\sigma}\alpha_{i}^{\left(1/2\right)\dagger}e_{i\uparrow}\nonumber \\
 & +\theta(\sigma)(U_{i}^{(e)\dagger}U_{i+1}^{(b)})_{\sigma,-\sigma}e_{i\uparrow}^{\dagger}\alpha_{i+1}^{(1/2)}+\textrm{H.c.}\},\label{68}
\end{align}
and
\begin{align}
\mathcal{H}^{J}= & -\frac{J}{4}\sum_{i;i'=i,i+1;\sigma}\theta(\sigma)|(U_{i}^{(d)\dagger}U_{i'}^{(b)})_{\sigma,\sigma}|^{2}\alpha_{i}^{\dagger}\alpha_{i}\nonumber \\
 & -\frac{J}{4}\sum_{i;i'=i,i+1;\sigma}\theta(\sigma)|(U_{i}^{(e)\dagger}U_{i'}^{(b)})_{\sigma,\sigma}|^{2}e_{i\uparrow}^{\dagger}e_{i\uparrow}\nonumber \\
 & -\frac{J}{4}\sum_{i;i'=i,i-1;\sigma}\theta(-\sigma)[|(U_{i}^{(b)\dagger}U_{i'}^{(d)})_{\sigma,\sigma}|^{2}\nonumber \\
 & +|(U_{i}^{(b)\dagger}U_{i'}^{(e)})_{\sigma,\sigma}|^{2}]\alpha_{i}^{(\frac{1}{2})\dagger}\alpha_{i}^{(\frac{1}{2})}.\label{69}
\end{align}
Notice that Eqs.~(\ref{68}) and (\ref{69}) are identical to Eqs.~(\ref{32c})
and (\ref{32d}), since Eqs.~(\ref{32a}) and (\ref{32b}) were eliminated
through the Legendre transformation. 

Some digression on $\mathcal{H}_{eff}^{t\textrm{-}J}$ is in order. One of the key properties of  
quasi-1D interacting quantum systems is the phenomenon of spin-charge
separation, leading to the formation of spin and charge-density waves,
which move independently and with different velocities. It has been
demonstrated~\cite{Montenegro2006} that for $\delta>2/3$ the low-energy
physics of the doped $AB_{2}$ Hubbard chain in the $U=\infty$ coupling
limit is described in terms of the Luttinger-liquid model, with the
spin and charge degrees of freedom decoupled. Most importantly, it has been shown that for the $AB_{2}$ $t$-$J$  Hubbard chains~\cite{Montenegro2014}
charge and spin quantum fluctuations are practically decoupled, as
suggested by the emergence of charge-density waves in anti-phase with
the modulation of the ferrimagnetic order. One can make use of this
feature to formally split each term of the $t$-$J$ Hamiltonian, Eq.~(\ref{eq:Htjeff})-(\ref{69}) ,
into a product of two independent terms acting on different Hilbert
spaces, i.e., we can enforce spin-charge separation and calculate
the charge and spin correlation functions in a decoupled fashion. 

Therefore, from the above  discussion, we  shall consider that the charge correlation functions are well described by an effective spinless tight-binding model~\cite{Montenegro2006,Oliveira_2009,Vitoriano2010}, since the hole (charge) density waves  develop along the $x$-axis and in anti-phase with the
modulation of the ferrimagnetic structure, as numerically observed in
Fig.~2(b) of Ref.~[\onlinecite{Montenegro2014}]. So, using   Eqs.~(\ref{GrassmannFieldsDirectSpace}), with $a/2\rightarrow a$ (effective lattice spacing of the linear chain: distance between $A$ and $B$ sites, see Fig.~2(a) of Ref.~[\onlinecite{Montenegro2014}]), we find
\begin{eqnarray}
 & \left\langle \right.\alpha_{i}^{\left(1/2\right)\dagger}\alpha_{i}\left.\right\rangle  & =\frac{1}{N_c}\sum_{kk'}e^{-ik(x_{i}-1)}e^{ik'x_{i}}\left\langle \varPsi_{0}\right.|\alpha_{k}^{\dagger}\alpha_{k'}|\left.\varPsi_{0}\right\rangle \nonumber \\
 &  & =\frac{1}{\pi}\int_{-k_{F}(\delta)}^{k_{F}(\delta)}e^{ik}dk=\frac{2}{\pi}\sin[k_{F}(\delta)],
\end{eqnarray} 
with $\left.|\varPsi_{0}\right\rangle $ being the hole-doped ferrimagnetic GS, where ${k_{F}(\delta)=\pi\frac{N_h}{N}\equiv\pi\delta}$ is the Fermi wave vector of the spinless tight-binding holes. In the same fashion: ${\left\langle \right.\alpha_{i}^{\dagger}\alpha_{i+1}^{\left(1/2\right)}\left.\right\rangle=\frac{2}{\pi}\sin[k_{F}(\delta)]}$ and ${\left\langle \right.\alpha_{i}^{\left(1/2\right)\dagger}e_{i\uparrow}\left.\right\rangle=\left\langle \right.e_{i\uparrow}^{\dagger}\alpha_{i+1}^{\left(1/2\right)}\left.\right\rangle=0}$; while $\left\langle \right.\alpha_{i}^{\dagger}\alpha_{i}\left.\right\rangle=\left\langle \right.\alpha_{i}^{\left(1/2\right)\dagger}\alpha_{i}^{\left(1/2\right)}\left.\right\rangle=\left\langle \right.e_{i\uparrow}^{\dagger}e_{i\uparrow}\left.\right\rangle=(1-\frac{1}{2\pi}\int_{-k_{F}(\delta)}^{k_{F}(\delta)}dk)=(1-\delta)$.  Here, we remark that the itinerant holes away from half filling are associated with the lower-energy dispersive $\alpha_{k}$ and $\alpha_{k}^{\left(1/2\right)}$ bands [see Fig.~\ref{Fig_1}\subref{Fig_1b} in Sec.~(\ref{sec:Functional})], thus contributing to the kinetic Hamiltonian in Eq.~(\ref{68}). On the other hand, the local correlations related to the lower-energy bands $\alpha_k$, $\alpha_{k}^{\left(1/2\right)}$, and $e_{k\uparrow}$, contribute equally to the exchange Hamiltonian in Eq.~(\ref{69}). Thereby, using the above tight-binding results for the charge correlation functions, $\mathcal{H}_{eff}^{t\textrm{-}J}$ in Eqs.~(\ref{eq:Htjeff})-(\ref{69}) gives rise to the $\delta$-dependent Hamiltonian, $\mathcal{H}_{eff}^{t\textrm{-}J}(\delta)=\mathcal{H}_{eff}^{t}(\delta)+\mathcal{H}_{eff}^{J}(\delta)$, written below: 
\begin{align}
\mathcal{H}_{eff}^{t\textrm{-}J}(\delta) & =-t\frac{2}{\pi}\sin[k_{F}(\delta)]\sum_{i}[(U_{i}^{(b)\dagger}U_{i}^{(d)})_{\downarrow\uparrow}\nonumber \\
 & +(U_{i}^{(d)\dagger}U_{i+1}^{(b)})_{\uparrow\downarrow}+\textrm{H.c.}]\nonumber \\
 & -\frac{J(1-\delta)}{4}\sum_{\left\langle i\alpha,j\beta\right\rangle \sigma}\theta(p_{i\alpha}\sigma)|(U_{i\alpha}^{\dagger}U_{j\beta})_{\sigma,\sigma}|^{2},\label{eq:htjspinsector}
\end{align}where the sum over $\sigma$ was evaluated in Eq.~(\ref{68}) and the square of the $SU(2)$ gauge field products in the exchange contribution have been summed up in Eq.~(\ref{69}), so that this contribution is just $(1-\delta)$ times $\mathcal{H}_{eff}^{J}$ at half filling, Eq.~(\ref{eq:Heisenberglike}), or alternatively,  in terms of spin fields, Eq.~(\ref{eq:heisenbergeff}), or spin-waves, Eqs.~(\ref{eq:HJeff})-(\ref{eq:SpinWaveDispersion}).  On the other hand, the $SU(2)$ gauge fields matrix
elements: $\left(U_{i}^{(b)\dagger}U_{i}^{(d)}\right)_{\downarrow\uparrow}$
and $\left(U_{i}^{(d)\dagger}U_{i+1}^{(b)}\right)_{\uparrow\downarrow}$,
that appear in the kinetic contribution  of Eq.~(\ref{eq:htjspinsector}), can be written in
terms of the spin fields~\cite{Oliveira_2009,Ribeiro2015} as
\begin{widetext}
\begin{equation}
(U_{i}^{(b)\dagger}U_{i}^{(d)})_{\downarrow\uparrow}+\textrm{H.c.}=\sum_{l}\frac{1}{\sqrt{2}}\left(\sqrt{1-2S_{i}^{B_{l},z}-2S_{i}^{A,z}+4S_{i}^{A,z}S_{i}^{B_{l},z}}+\sqrt{1+2S_{i}^{A,z}+2S_{i}^{B_{l},z}+4S_{i}^{A,z}S_{i}^{B_{l},z}}\right),\label{eq:Ubdi}
\end{equation}
and $\left(U_{i}^{(d)\dagger}U_{i+1}^{(b)}\right)_{\uparrow\downarrow}$ is obtained from Eq.~(\ref{eq:Ubdi}) through the replacement  $S_{i}^{A,z}\rightarrow S_{i+1}^{A,z}$, in which case we took advantage of the $U(1)$ gauge freedom and Eq.~(\ref{Spialpha}). Notice that these square-root matrix elements depend on $z$-spin components only. 

At this stage, it will prove useful, in the calculation of the GS total spin in the doped regime, to consider $\mathcal{H}_{eff}^{t\textrm{-}J}(\delta,h)$
which describes the system in the presence of a homogeneous magnetic
field $\mathbf{h}=h\mathbf{\hat{z}}=(-h_{A}+h_{B_{1}}+h_{B_{2}})\mathbf{\hat{z}}$,
where the staggered fields point along the local corresponding magnetizations
in the ferrimagnetic phase have the same magnitude $h$. The magnetic field couple with
the spin fields through the Zeeman term (see Sec.~\ref{Sublattice-magnetizations}) \textit{and} with
the charge degrees of freedom through the magnetic orbital coupling
in the Landau gauge: $\mathbf{A}=hx\mathbf{\hat{y}}$. Since our aim is to study doping effect on the magnetization, we shall assume vanishingly small magnetic field in the context of linear response theory and  perturbative expansion in the strong-coupling regime. Additionally, the magnetic
orbital coupling can be considered through the so-called Peierls substitution~\cite{Essler2005,Vidal2000,*Gulacsi2007}: $t\rightarrow te^{i\int_{_{i\alpha}}^{_{j\beta}}\mathbf{A}\cdot d\mathbf{l}}$,
where $i\alpha$ and $j\beta$ are first-neighbor sites, and the flux quantum $\phi_{0}=hc/e\equiv1$. If one consider that  the carrier is at the site $iA$, we
have four hopping possibilities: $iA\rightarrow iB_{1,2}$ and $iA\rightarrow(i+1)B_{1,2}$, so the total phase $\phi$ acquired by the carrier in this prescription
satisfies Stokes' theorem: $\phi=\ointctrclockwise_{\mathrm{unit\,cell}}\mathbf{A}\cdot d\mathbf{l}=\iint_{S}\mathbf{h}\cdot d\mathbf{S}=ha^{2}$
$(a\equiv1)$. We also remark that, in order to obtain real values for the zero-field staggered magnetizations,
we have considered, for convenience, an imaginary gauge transformation~\cite{Ribeiro2015,Nelson1996}:
$\mathbf{A}\rightarrow i\mathbf{A}$.  Therefore, by placing Eq.~(\ref{eq:Ubdi}) and the similar matrix element into the kinetic term in Eq.~(\ref{eq:htjspinsector}), making the above Peierls substitution, and using the Holstein-Primakoff and Bogoliubov transformations introduced in Eqs.~(\ref{eq:Holstein-PrimakoffA})-(\ref{expansion}) and Eqs.~(\ref{eq:Bogoliubov transformation})-(\ref{BogoliubovUVk}), respectively, up to order $\mathcal{O}(S^{-1})$, we arrive at the following diagonalized  kinetic Hamiltonian $\mathcal{H}_{eff}^{t}(\delta,h)$:
\begin{equation}
\mathcal{H}_{eff}^{t}(\delta,h)=-\frac{4\sqrt{2}}{\pi}te^{-(-h_{A}+h_{B_{1}}+h_{B_{2}})}\sin[k_{F}(\delta)]\sum_{k}[4S-3v_{k}^{2}-(u_{k}^{2}+2v_{k}^{2})\alpha_{k}^{\dagger}\alpha_{k}-(2u_{k}^{2}+v_{k}^{2})\beta_{k}^{\dagger}\beta_{k}-\xi_{k}^{\dagger}\xi_{k}],\label{eq:Hteff}
\end{equation}where the doped-induced contributions for the spin dispersion relations are evidenced in the last three terms. On the other hand, by adding  the Zeeman terms (see Sec.~\ref{Sublattice-magnetizations}) to the  exchange contribution $\mathcal{H}_{eff}^{J}(\delta)$, given in Eq.~(\ref{eq:htjspinsector}), we obtain $\mathcal{H}_{eff}^{J}(\delta,h)$. Lastly, by adding  the kinetic and the exchange contributions, we arrive at the effective $t$-$J$ Hamiltonian in the presence  of a magnetic field:

\begin{eqnarray}
\mathcal{H}_{eff}^{t\textrm{-}J}(\delta,h) & = & -\frac{4\sqrt{2}}{\pi}te^{-(-h_{A}+h_{B_{1}}+h_{B_{2}})}\sin[k_{F}(\delta)]\sum_{k}(4S-3v_{k}^{2})+J(1-\delta)(E_{GS}^{J}-JN_{c})\nonumber \\
 &  & +\sum_{k}[\epsilon_{k}^{(\alpha)}(\delta)\alpha_{k}^{\dagger}\alpha_{k}+\epsilon_{k}^{(\beta)}(\delta)\beta_{k}^{\dagger}\beta_{k}+\epsilon_{k}^{(\xi)}(\delta)\xi_{k}^{\dagger}\xi_{k}]-h_{A}\sum_{i}S_{i}^{A,z}-h_{B_{1}}\sum_{i}S_{i}^{B_{1},z}-h_{B_{2}}\sum_{i}S_{i}^{B_{2},z},\label{Htotal}
\end{eqnarray}
where $E_{GS}^{J}$ is given by Eq.~(\ref{eq:EJGS}) and (\ref{eq:EGSN}), and the corresponding spin-wave modes [see Eqs.~(\ref{eq:Hteff}), (\ref{eq:espinwave}), (\ref{dispersioncorrection2})-(\ref{eq:q1q2}), and (\ref{eq:SpinWaveDispersion})] of the doped
$AB_{2}$ $t$-$J$ chain read:
\begin{equation}
\epsilon_{k}^{(\alpha)}(\delta)  =\frac{4\sqrt{2}}{\pi}t\sin(\pi\delta)[u_{k}^{2}+2v_{k}^{2}]+(1-\delta)(\epsilon_{k}^{0(\alpha)}+\delta\epsilon_{k}^{(\alpha)}),\label{spinwavedelta1}
\end{equation}
$\epsilon_{k}^{(\beta)}(\delta)$ is obtained from $\epsilon_{k}^{(\alpha)}(\delta)$
through the exchange $u_{k}\leftrightarrow v_{k}$ and the replacement $\alpha\rightarrow \beta$, while
\begin{equation}
\epsilon_{k}^{(\xi)}(\delta)=\frac{4\sqrt{2}}{\pi}t\sin(\pi\delta)+(1-\delta) (\epsilon_{k}^{0(\xi)}+\delta\epsilon_{k}^{(\xi)}).\label{spinwavedelta3}
\end{equation}
We find it instructive to comment on the analytical structure of the above equations. Firstly, we mention the presence of the Bogoliubov parameters [see Eqs.~(\ref{BogoliubovUVk})] in a symmetric form in the kinetic terms 
of Eq.~(\ref{spinwavedelta1}) and its analogous for $\epsilon_{k}^{(\beta)}(\delta)$; besides,
although the flat mode is strongly affected by the presence of holes,
it remains dispersionless. In addition, using Eqs.~(\ref{Htotal}) and (\ref{eq:q1q2}), the total GS energy (no spin-wave excitations) per unit cell in the thermodynamic limit is readily obtained:
\begin{alignat}{1}
E_{GS}^{t\textrm{-}J}(\delta,h)/N_{c} & =-\frac{4\sqrt{2}}{\pi}te^{-(-h_{A}+h_{B_{1}}+h_{B_{2}})}\sin(\pi\delta)(4S-3q_{1})\nonumber \\
 & +(1-\delta)(E_{GS}^{J}/N_{c}-J)-\left\langle \right.S^{A,z}\left.\right\rangle h_{A}-\sum_{l=1,2}\left\langle \right.S^{B_{l},z}\left.\right\rangle h_{B_{l}},\label{EGStJ}
\end{alignat}where  $\left\langle \right.S^{A,z}\left.\right\rangle$ and $\left\langle \right.S^{B_{l},z}\left.\right\rangle$ are the calculated  sublattice magnetizations, at half filling and zero-field,  given by Eqs.~(\ref{SA}). 
\end{widetext}

In subsections~\ref{subsec:Spin-Wave}, \ref{subsec:t-JA}, and \ref{subsec:t-JB},  we will show that the underlying competing physical mechanisms: the magnetic orbital response and the Zeeman contribution embedded in Eqs.~(\ref{Htotal})-(\ref{EGStJ}) will dramatically affect the behavior of the system under hole doping and, in particular, will lead to spiral IC spin structures, the breakdown of the spiral ferrimagnetic GS at a critical value of the hole doping, a region of phase separation, and RVB states at $\delta\approx1/3$.

\subsection{\label{subsec:Spin-Wave}Doped regime: Spin-wave modes}
Before we go one step further to discuss relevant macroscopic quantities, i.e., the GS energy and total spin in the doped regime, we shall first undertake a detailed study, at a microscopic level, of the hole-doping effect on the calculated spin-wave branches given by Eqs.~(\ref{spinwavedelta1})-(\ref{spinwavedelta3}).

\begin{figure}
\centering
\subfloat{\label{Fig_2a}\includegraphics[scale=1]{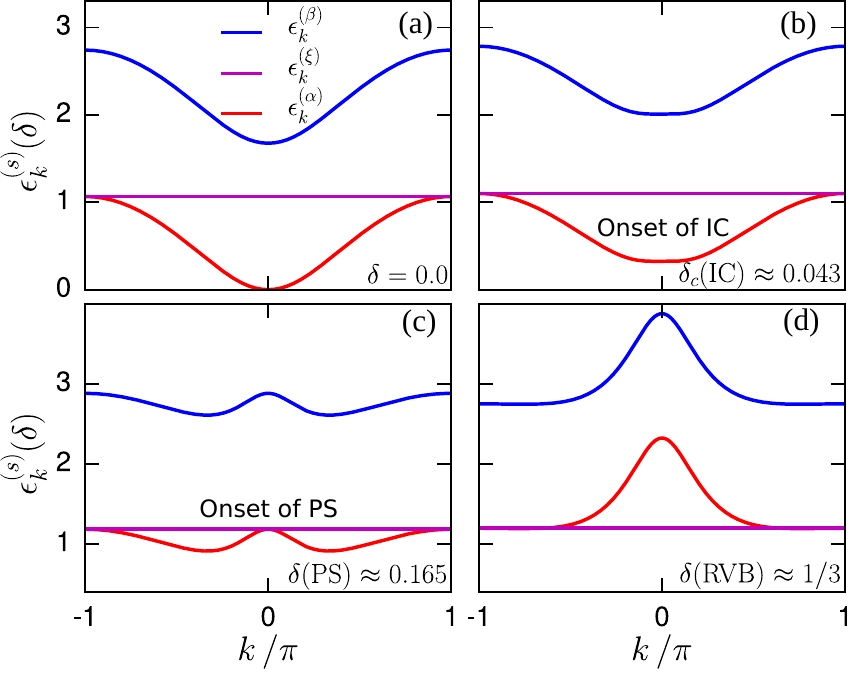}}
\subfloat{\label{Fig_2b}}
\subfloat{\label{Fig_2c}}
\subfloat{\label{Fig_2d}}
\caption{(Color online) Evolution of the zero-field second-order spin-wave
dispersion relations of the $AB_{2}$ $t\textrm{-}J$ chain
as a function of hole doping ($\delta$):   dispersive ferromagnetic  $\epsilon_{k}^{(\alpha)}$ and antiferromagnetic  $\epsilon_{k}^{(\beta)}$ modes and the flat ferromagnetic one  $\epsilon_{k}^{(\xi)}$,  at (a) half filling;  (b)  the  onset  of the spiral IC spin structures at   $\delta_{c}(\mathrm{IC})=0.043$, in which case the flattening  of the gap of   $\epsilon_{k}^{(\alpha)}$ around  $k=0$ is observed;  (c)  the onset of PS at $\delta(\mathrm{PS})=0.165$, characterized by the overlap of the two ferromagnetic modes at $k = 0$ and by the spatial coexistence of two phases: spiral IC spin structures, with modulation fixed at $\delta(\mathrm{PS})$, and RVB states at $\delta\approx1/3$. (d) At $\delta=1/3$  the flat mode presents the lowest energy, thus indicating that the short-range RVB state is the stable phase.}
\label{Fig_2}
\end{figure}

Fig.~\ref{Fig_2} depicts the second-order spin-wave dispersion relations at $J/t=0.3$ and for the indicated values of $\delta$. Without loss of generality, we set $t=1$ in our numerical computations.  At half filling, the antiferromagnetic mode $\epsilon_{k}^{(\beta)}$, together with the two ferromagnetic modes: the dispersive $\epsilon_{k}^{(\alpha)}$ and the flat one $\epsilon_{k}^{(\xi)}$, are shown in Fig.~\ref{Fig_2}\subref{Fig_2a},  which are defined in Eq.~(\ref{eq:SpinWaveDispersion}), and can be plotted using Eqs.~(\ref{eq:espinwave}) and (\ref{dispersioncorrection2})-(\ref{eq:q1q2}).

As the hole doping increases slightly, the abrupt decrease of the peaks at $k=0$ and $k=\pi$ of the  numerical DRMG structure factor (see Fig.~3 of Ref.~[\onlinecite{Montenegro2014}]), associated with the ferrimagnetic order, manifests itself here through the opening of a gap in the ferromagnetic Goldstone mode $\epsilon_{k}^{(\alpha)}$, as seen in Fig.~\ref{Fig_2}\subref{Fig_2b},  thus indicating that the system loses its long-range order. Note that the antiferromagnetic mode $\epsilon_{k}^{(\beta)}$ is also similarly shifted. On the other hand, although the dispersion relation is modified for small values of the wave vector $k$, the minimum value of $\epsilon_{k}^{(\alpha)}$ still remains at $k=0$ up to the \textit{onset} of the formation of spiral IC spin structures at $\delta_{c}(\mathrm{IC})=0.043$ (a value that should be compared with the numerical DMRG estimate of $\delta\approx0.055\pm0.012$), characterized by the flattening of the dispersive spin-wave branches around zero. Upon further increase of $\delta$, two minima form (around $k=0$) and  move away from each other as one enhances the hole doping. This behavior is the signature of the occurrence of spiral IC spin structures (see Fig.~3 of  Ref.~[\onlinecite{Montenegro2014}]).

Fig.~\ref{Fig_2}\subref{Fig_2c} shows the onset of phase separation (PS) at $\delta(\mathrm{PS})=0.165$ for $J=0.3$, which is characterized by the overlap  of the two ferromagnetic modes at $k=0$. The signature of this regime is the spatial coexistence of two phases: spiral IC spin structures at $\delta(\mathrm{PS})=0.165$ and RVB states at $\delta\approx1/3$, in very good agreement with the numerical estimate of $\delta_{\mathrm{IC-PS}}\approx0.16$~\cite{Montenegro2014}. At  $\delta\approx1/3$, the flat mode has the lowest energy, as illustrated in Fig.~\ref{Fig_2}\subref{Fig_2d}. This behavior indicates that the RVB state is the stable phase at $\delta\approx1/3$ and $J=0.3$~\cite{Montenegro2014}, and also in agreement with the numerical DMRG studies~\cite{Sierra1999,Montenegro2006} and analytical prediction at $U=\infty$~\cite{Oliveira_2009}.

\begin{figure}
\includegraphics[scale=1]{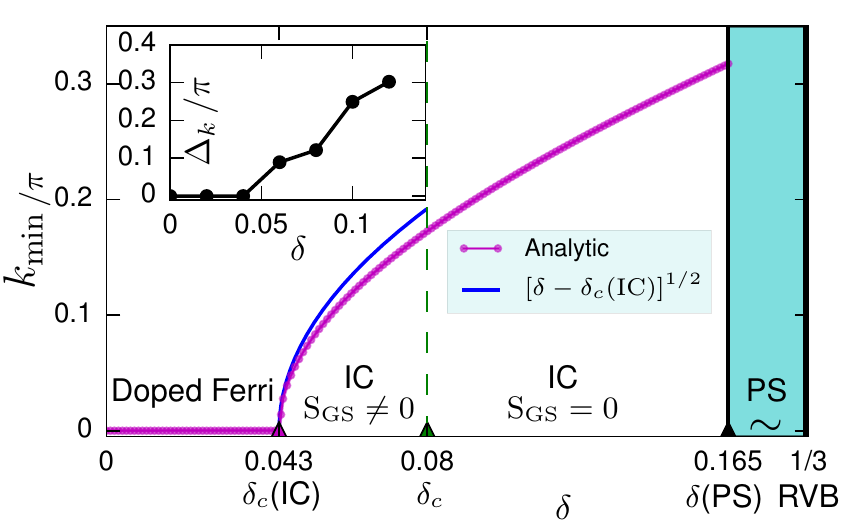}\caption{\label{Fig_3}(Color online) Evolution of $k_{\mathrm{min}}$ (value
of $k$ at the local minimum of $\epsilon_{k}^{(\alpha)}(\delta)$ near $k=0$) as a function of $\delta$:  doped ferrimagnetism for $0<\delta<\delta_c(\mathrm{IC})\approx0.043$; spiral IC spin structures with non-zero (zero)  $S_{GS}$ for  $\delta_c(\mathrm{IC})<\delta<\delta_c\approx0.08$ ($\delta_c<\delta<\delta(\mathrm{PS})\approx0.165$),  with a second-order quantum phase transition at $\delta_c(\mathrm{IC})$ characterized  by a  square-root  behavior  $[\delta-\delta_c(\mathrm{IC})]^{1/2}$ (blue line),  and a first-order transition at $\delta(\mathrm{PS})$ involving the IC spin structure, with modulation fixed at $\delta(\mathrm{PS})$, and short-range RVB states at hole concentration 1/3.   The  inset shows DMRG data  from  Ref.~[\onlinecite{Montenegro2014}] for   $\Delta_{k}\equiv k_{\mathrm{max}}-\pi$ as a function of $\delta$, where $k_{\mathrm{max}}$ is the value of $k$ at  the local maximum of the structure factor $S(k)$ near $k=\pi$, in qualitative agreement with the second-order transition  at $\delta_c(\mathrm{IC})$.}
\end{figure}

In order to better understand the rich variety of doping-induced phases in the system,  in Fig.~\ref{Fig_3} we plot the evolution of the wave vector $k_{\mathrm{min}}$ corresponding to the local minimum of $\epsilon_{k}^{(\alpha)}(\delta)$, upon increasing the hole doping $\delta$ from $0$ to $1/3$. The wave vector $k_{\mathrm{min}}$   remains zero until it hits the onset doping value $\delta_c(\mathrm{IC})=0.043$, beyond which a square-root growth behavior takes place~\cite{Chakrabarty2012}: $[\delta-\delta_c(\mathrm{IC})]^{1/2}$ (blue line), for $\delta$ close to $\delta_c(\mathrm{IC})$. The square-root growth behavior is the signature of the occurrence of a second-order quantum phase transition from the doped ferrimagnetic phase to the IC spiral ferrimagnetic state with a non-zero value of the total GS spin, $S_{GS}$. This result is supported by  the behavior of $\Delta_{k}\equiv k_{\mathrm{max}}-\pi$ at which the local maximum of the numeric DMRG structure factor $S(k)$ near $k=\pi$ is observed, as shown  in the inset of Fig.~\ref{Fig_3} (taken from  the inset of Fig.~3(b) of Ref.~[\onlinecite{Montenegro2014}]). 
For further increase of hole doping our result deviates from the square-root
growth behavior and some very interesting features are to be noticed.
The value of $\delta_c=0.08$ indicates the breakdown of the total $S_{GS}$
in the IC phase, as will be confirmed by the explicit calculation
of $S_{GS}$, a macroscopic quantity, in Section~\ref{subsec:t-JB}. Thus, for $0.08<\delta<0.165$
the system displays an IC phase with zero $S_{GS}$, in agreement with
the DMRG data (see Fig.~1(c) of Ref.~[\onlinecite{Montenegro2014}]). At $\delta(\mathrm{PS})=0.165$ the system exhibits a first-order transition accompanied by the spatial phase separation regime:   the IC phase with zero $S_{GS}$ and modulation
fixed by $\delta(\mathrm{PS})$ in coexistence with  the short-range RVB states at
$\delta\approx1/3$, also consistent with the DMRG data plotted in Fig.~4 of  Ref.~[\onlinecite{Montenegro2014}].

Lastly, we emphasize that, despite the occurrence of several  doping-induced phases in the DMRG studies~\cite{Montenegro2014}: Lieb ferrimagnetism, spiral IC spin structures,  RVB states with finite spin gap, phase separation, and Luttinger-liquid behavior, it is surprising and very interesting that the second-order spin-wave modes remain stable up to $\delta\approx1/3$, with predictions in very good agreement with the DMRG studies~\cite{Montenegro2014}.  In this context, it is worth mentioning the long time studied case of rare earth metals~\cite{Elliott1963,*[See also: ][{, Chapter 1, p. 187.}]Elliott1972magnetic}, where an external magnetic field can induce non-trivial  phase transitions involving spiral spin structures, well described by spin-wave theory.

\subsection{\label{subsec:t-JA}Doped regime: Ground state energy}

Performing the integration over the first BZ  in Eqs.~(\ref{eq:q1q2}) and (\ref{eq:EGSN})
and setting $S=1/2$ in Eq.~(\ref{EGStJ}), we find that the $AB_{2}$ $t$-$J$ ground
state energy per unit cell as a function of hole doping in zero-field
reads:
\begin{equation}
E_{GS}^{t\textrm{-}J}(\delta)/JN_{c} =-1.9543\frac{t}{J}\sin\left(\pi\delta\right)-2.4608\left(1-\delta\right).\label{eq:ENERGYTJ}
\end{equation}

We shall now examine the case of small hole doping away from half
filling, i.e., with  hole concentration ranging from $\delta=0$
up to $\delta=0.2$ for two values of  $J$: $0.1$ and
$0.3$.
\begin{figure}
\includegraphics[scale=1]{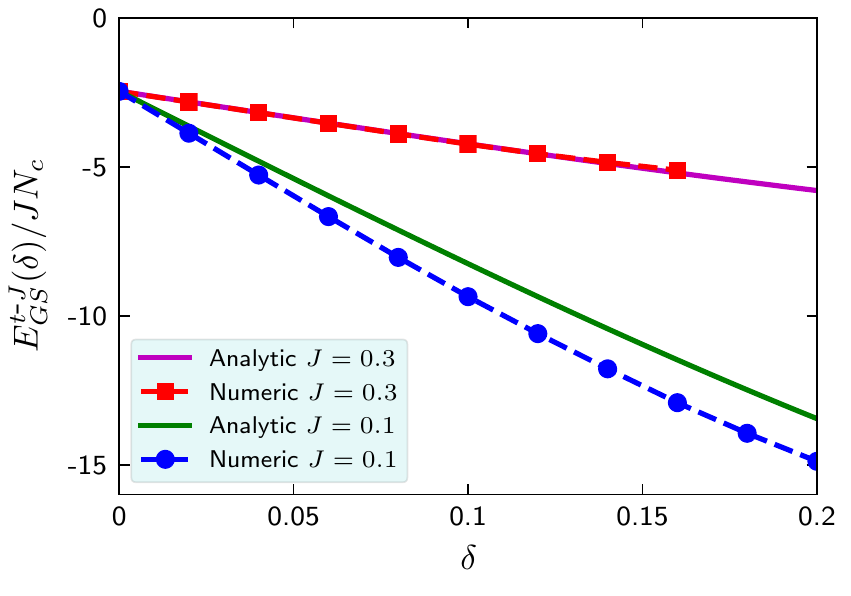}\caption{\label{Fig_4}(Color online) Analytical prediction
for the GS energy per unit cell of the $AB_{2}$ $t\textrm{-}J$
chain as a function of doping, and comparison with numerical data
from DMRG  technique for $J/t=0.1$ and $J/t=0.3$~\cite{Montenegro2014}.
At half filling ($\delta=0$), both results meet at the  expected prediction~\cite{Montenegro2014}: $\approx-2.4678$.
Note that we have added the term $-JN_{c}$ with the intention of comparison with numerical calculation.}
\end{figure}
In Fig.~\ref{Fig_4}, we show the evolution of
the GS energy per unit cell of the $AB_{2}$ $t$-$J$ model as a
function of hole doping for both mentioned values of $J$, and the  
comparison was made with the numerical DMRG data~\cite{Montenegro2014}. From
the two results at $J=0.3$, the only quantitative
difference induced by the increase of the hole concentration is a
crossing feature around $\delta\approx0.1$,  where our analytical
result slightly change its behavior by lowering the energy with respect
to the numerical data~\cite{Montenegro2014}. In fact, because  our model assumes a ferrimagnetic state as the starting point, this change of behavior suggests that we have entered in a region of strong magnetic instabilities,
and possibly indicating a smooth transition to an incommensurate phase
with zero GS total spin beyond $\delta\approx0.1$, as confirmed by
the numerical data in Ref.~[\onlinecite{Montenegro2014}] and illustrated in  Fig.~\ref{Fig_3}. On the
other hand, at $J=0.1$, although our results reproduce the numerical
data with an acceptable agreement, we observe a discrepancy that
increases with $\delta$. The cause of such discrepancy will be discussed
in the next subsection.

\begin{figure}
\centering
\subfloat{\label{Fig_5a}\includegraphics[scale=1]{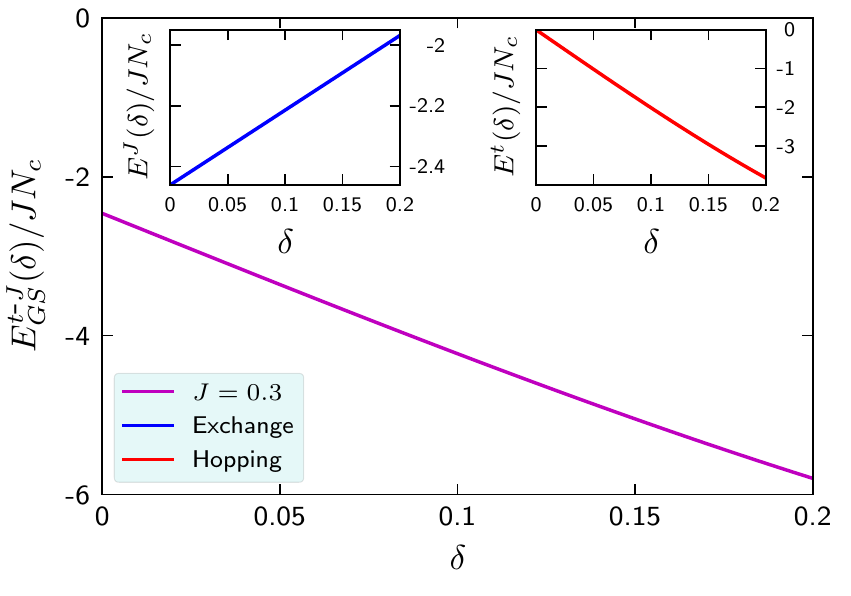}}
\subfloat{\label{Fig_5b}}
\caption{(Color online) Ground-state energy per unit cell for the $AB_{2}$
$t\textrm{-}J$ chain as a function of $\delta$ for $J=0.3$.
In the insets, we illustrate the two energetic contribution due to
(a) exchange and (b) hopping terms.}
\label{Fig_5}
\end{figure}

With the purpose of determining the interplay between the contribution
of magnetic exchange and the itinerant kinetic energy  to the zero-field GS energy Eq.~(\ref{EGStJ}),
we take $J=0.3$ and show its evolution with doping in Fig.~\ref{Fig_5}.
We can see in the insets, Fig.~\ref{Fig_5}\subref{Fig_5a}
and Fig.~\ref{Fig_5}\subref{Fig_2b},
the competitive behavior of the two energetic contributions, i.e.,
the contribution of the exchange energy increases linearly  with $\delta$,
while a practically linear decrease of the hopping term is observed
as one enhances the hole doping. This competition indicates that a
phase transition to a paramagnetic phase should occur at some critical
concentration value.

\subsection{\label{subsec:t-JB}Doped regime: Ground state total spin }
The existence of a transition from an IC spiral ferrimagnetic
phase to an IC paramagnetic one is a most interesting
feature observed numerically in doped $AB_{2}$ $t$-$J$ Hubbard
chains~\cite{Montenegro2014}. In order to firmly corroborate the mentioned transition, we have calculated the GS total spin per unit cell as a function of hole doping, $S_{GS}(\delta)=\sum_{\alpha}\,\left\langle \right.S^{\alpha,z}\left.\right\rangle(\delta)$, with $\alpha=A,B_{1},B_{2}$, by means of the zero-field derivative of Eq.~(\ref{EGStJ}):\begin{equation}
\left\langle \right.S^{\alpha,z}\left.\right\rangle(\delta)=-(1/N_{c})[\partial E_{GS}^{t\textrm{-}J}(\delta,h)/\partial h_{\alpha}]|_{h_{\alpha}=0}.
\end{equation}We thus  find\begin{equation}
\left\langle \right.S^{\alpha,z}\left.\right\rangle(\delta)=\left\langle \right.S^{\alpha,z}\left.\right\rangle\pm\frac{4\sqrt{2}}{\pi}t\sin(\pi\delta)(4S-3q_{1}),\label{Salpha}
\end{equation}
where  $+$ ($-$) corresponds to  sublattice $\alpha=A$ ($\alpha=B_{1,2}$), and  $\left\langle \right.S^{A,z}\left.\right\rangle$ and $\left\langle \right.S^{B_{l},z}\left.\right\rangle$ are given by Eqs.~(\ref{SA}). Therefore, by performing the integration over the first BZ of the three contributions in Eq.~(\ref{Salpha}), we finally obtain:\begin{equation}
\frac{S_{GS}\left(\delta\right)}{S_{L}}=1-3.9086\sin(\pi\delta),\label{eq:87}
\end{equation}
where $S_L=\sum_{\alpha}\,\left\langle \right.S^{\alpha,z}\left.\right\rangle=1/2$ is  Lieb's reference value for the GS total spin per unit cell at half filling and zero-field (see Section~\ref{Sublattice-magnetizations}).

\begin{figure}
\includegraphics[scale=1]{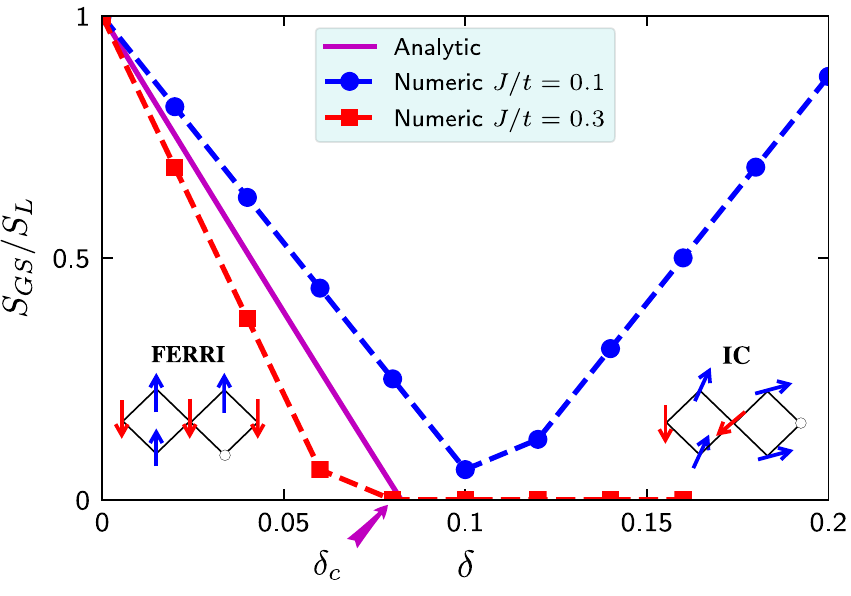}
\caption{\label{Fig_6} (Color online) Ground-state total spin $S_{GS}$ per unit cell (solid magenta line), normalized by its
value in the undoped regime: $S_{L}=\frac{1}{2}$, as a function of
hole doping $\delta$ for the indicated values of $J$. In the figure, $\delta_c\approx0.08$ indicates  the critical value of doping at which the magnetic order is suppressed  and a second-order phase transition takes place.}
\end{figure}

In Fig.~\ref{Fig_6} we plot the evolution
of $S_{GS}$, normalized by $S_{L}$, as a function of $\delta$, and 
compare it with the numerical data from DMRG and Lanczos techniques~\cite{Montenegro2014},
for $J=0.3$ (red squares) and $J=0.1$ (blue circles). In the
latter (former) case, the system undergoes a transition from the modulated  itinerant
ferrimagnetic phase to an incommensurate phase with zero (nonzero) $S_{GS}$.  Notice that, in both cases, the transition is characterized by  a  decrease of  $S_{GS}$  from $S_{L}$ to $0$ or to a residual value, 
regardless of the value that $S_{GS}$  takes after the transition. Indeed, at $J=0.1$ and $\delta>0.1$, the formation of magnetic polarons (onset of the Nagaoka
phenomena that sets in as $U\rightarrow\infty$) with charge-density
waves in phase with the modulation of the ferrimagnetic structure,
as indicated by the DMRG data~\cite{Montenegro2014},  leads to
an incommensurate phase with nonzero $S_{GS}$.

Most importantly, we can observe in Fig.~\ref{Fig_6} that the value of $S_{GS}$ decreases practically linearly with $\delta$ until the magnetic order is completely suppressed at $\delta_{c}\approx0.08$. This behavior is supported by numerical results~\cite{Montenegro2014}, particularly in the regime where the Nagaoka phenomenon is not manifested, that is, at
$J=0.3$,  as  indicated in Fig.~\ref{Fig_3}. In this regime, spin and charge quantum fluctuations destabilize the ferrimagnetic structure and trigger a transition to an incommensurate paramagnetic phase at $\delta_{c}$, with $S_{GS}\sim(\delta-\delta_{c})\rightarrow0$.

\section{\label{sec:CONCLUSIONS}CONCLUSIONS}
In summary, we have presented a detailed analytical study of the large-U
Hubbard model on the quasi-one-dimensional $AB_{2}$ chain. We used
a functional integral approach combined with a perturbative expansion
in the strong-coupling regime that allowed us  to properly analyze the
referred system at and away from half filling. 

At half filling, our model was mapped onto the quantum Heisenberg model, and analyzed through a spin-wave perturbative series expansion in powers of $1/S$. We have demonstrated that the GS energy, spin-wave
modes, and sublattice magnetizations are in very good agreement with previous results.
In the challenging hole doping regime away from half filling, the corresponding $t\textrm{-}J(=4t^{2}/U)$ Hamiltonian was derived. Further, under the assumption that charge and spin quantum correlations are decoupled,  the evolution of the second-order spin-wave modes in the doped regime has unveiled the occurrence of spatially modulated spin structures and the emergence of phase separation  (first-order transition)  in the presence of resonating-valence-bond states. The doping-dependent GS energy and total spin per unit cell are also calculated, in which case the  collapse  of the spiral magnetic order at a critical hole concentration was observed. Remarkably, our above-mentioned analytical results in the doped regime  are in very good agreement with density matrix renormalization group  studies, where our assumption of spin-charge decoupling is numerically supported by the formation of charge-density waves in anti-phase with the modulation of the ferrimagnetic structure.

Finally, we stress that our reported results evidenced that the present approach, also used in a study on the compatibility between numerical and analytical outcomes of the large-U Hubbard model on the honeycomb lattice, was proved suitable for the $AB_{2}$ chain (a quasi-1D system), where the impact of charge and spin quantum fluctuations are expected to manifest in a stronger way. We thus conclude that our approach offers a quite powerful analytical description of hole-doping induced phases away from half filling in low-dimensional strongly-correlated electron systems.
\begin{acknowledgments}
We appreciate interesting discussions with R. R. Montenegro-Filho.
This work was supported by CNPq, CAPES and FACEPE/PRONEX (Brazilian agencies).
\end{acknowledgments}

\bibliography{apssamp}

\end{document}